\documentclass[aps,pra,epsf,superscriptaddress,amsmath,amssymb,amsfonts,twocolumn,showpacs]{revtex4-1}

\usepackage{graphicx}
\usepackage{epsfig}
\usepackage{dcolumn}
\usepackage{bm}
\usepackage{braket}
\usepackage{amsmath}
\usepackage{mathtools}
\usepackage{graphicx,color,xcolor}
\usepackage{hyperref}
\newcommand{\abs}[1]{\left| #1 \right|} 
\usepackage{hyperref}
\usepackage{multirow}

\usepackage[normalem]{ulem} 

\begin{document}
\title{Formation and quench of homonuclear and heteronuclear quantum droplets in one dimension}

\author{S. I. Mistakidis}
\affiliation{Center for Optical Quantum Technologies, Department of Physics, University of Hamburg, 
Luruper Chaussee 149, 22761 Hamburg Germany}
\affiliation{ITAMP, Center for Astrophysics~$|$~Harvard $\&$ Smithsonian, Cambridge, MA 02138 USA} 
\author{T. Mithun}
 \affiliation{Department of Mathematics and Statistics, University of Massachusetts, Amherst MA 01003-4515, USA}
\author{P. G. Kevrekidis}
 \affiliation{Department of Mathematics and Statistics, University of Massachusetts, Amherst MA 01003-4515, USA}
\author{H. R. Sadeghpour}
\affiliation{ITAMP, Center for Astrophysics~$|$~Harvard $\&$ Smithsonian, Cambridge, MA 02138 USA} 
\author{P. Schmelcher}
\affiliation{Center for Optical Quantum Technologies, Department of Physics, University of Hamburg, Luruper Chaussee 149, 22761 Hamburg Germany} 
\affiliation{The Hamburg Centre for Ultrafast Imaging,
University of Hamburg, Luruper Chaussee 149, 22761 Hamburg, Germany}

\date{\today}

\begin{abstract} 
We exemplify the impact of beyond Lee-Huang-Yang (LHY) physics, especially due to intercomponent correlations, in the ground state and the quench dynamics of one-dimensional so-called quantum droplets using an ab-initio nonperturbative approach. 
It is found that the droplet Gaussian-shaped configuration arising for intercomponent attractive couplings
becomes narrower for stronger intracomponent repulsion and transits towards a flat-top structure either for larger particle numbers or weaker intercomponent attraction. 
Additionally, a harmonic trap prevents the flat-top formation.  
At the balance point where mean-field interactions cancel out,  we show that a correlation hole is present in the few particle limit of these fluids as well as for flat-top droplets. 
Introducing mass-imbalance, droplets experience intercomponent mixing and excitation signatures are identified for larger masses. 
Monitoring the droplet expansion (breathing motion) upon considering interaction quenches to stronger (weaker) attractions we explicate that beyond LHY correlations 
result in a reduced velocity (breathing frequency). 
Strikingly, the droplets feature two-body anti-correlations (correlations) at the same position (longer distances). 
Our findings pave the way for probing correlation-induced phenomena of droplet dynamics in current ultracold atom experiments. 
\end{abstract}

\maketitle

\section{Introduction} 

Quantum droplets, self-bound many-body states of matter which emerge when strong attraction is present, are a prominent manifestation of beyond mean-field (MF) effects among ultradilute weakly interacting systems~\cite{papp2008bragg,navon2010equation,navon2011dynamics,shin2008realization}. They form in the presence of quantum fluctuations~\cite{petrov2015quantum,petrov2016ultradilute} which stabilize an atomic gas, otherwise prone, per its
mean-field interactions, to collapse. 
Droplets have seen recent experimental demonstrations in homonuclear~\cite{cabrera2018quantum,cheiney2018bright,ferioli2019collisions} and heteronunclear~\cite{d2019observation,burchianti2020dual} two and three-dimensional ultracold bosonic mixtures as well as in dipolar condensates~\cite{ferrier2016observation,chomaz2016quantum} where signatures of supersolidity have been reported. On the theoretical side,
for two-component Bose mixtures with zero-range intracomponent repulsion and intercomponent attraction, droplets 
realized in models with next to leading order Lee-Huang-Yang (LHY)~\cite{huang1957quantum} corrections, have been described by a modified Gross-Pitaevskii (MGP) equation~\cite{petrov2015quantum,petrov2016ultradilute}; see also the recent review of Ref.~\cite{luo2021new}. 
Quantum droplets have been intensively studied in two and three dimensions for their topological properties in the presence of vortices~\cite{luo2021new,kartashov2018three}. 
Their self evaporation due to lossy inelastic collisions~\cite{ferioli2019collisions,cabrera2018quantum} that prevent droplet observation at long timescales, has also been considered. 
This is a process that can be delayed using heteronuclear mixtures~\cite{d2019observation,fort2021self} or supressed using one-dimensional (1D) settings~\cite{tolra2004observation,lavoine2021beyond}. These recent developments clearly underscore the
relevance of such structures in a wide range of theoretical
and experimental studies and their interest in quantifying the
interplay between mean-field and quantum effects.

In this vein, it  is of  particular interest
to examine the validity of the MGP framework in different interaction regimes using either higher-order quantum corrections~\cite{ota2020beyond,gu2020phonon,hu2020microscopic} or nonperturbative approaches~\cite{cikojevic2019universality,parisi2020quantum,parisi2019liquid}. 
1D settings constitute promising platforms for probing beyond-LHY physics since this is where strongly correlated regimes can be easily reached~\cite{kinoshita2006quantum,li2020relaxation}. A particularly relevant feature that has been an explicit
target of experimental~\cite{cheiney2018bright} and theoretical~\cite{luo2021new} studies is the droplet ground state. 
The latter in homogeneous bosonic mixtures is known to exhibit a crossover from a flat-top (FT) to a Gaussian-shaped configuration avoiding collapse due to quantum fluctuations despite the strong attractive interactions~\cite{astrakharchik2018dynamics}. 

Recently there has been a number of further developments
regarding the physics of quantum droplets. More specifically, the
excitation of the droplets has been examined~\cite{tylutki2020collective} and it has been found that the breathing mode therein is always a bound state independently
of the interaction. 
Moreover, thermal instabilities driving the liquid to gas transition in the evaporation of 1D liquids have been analyzed~\cite{de2021thermal,wang2020thermal}. 
Quantum Monte-Carlo approaches~\cite{parisi2020quantum,parisi2019liquid}
revealed that only close to the balance point of MF repulsion and attraction beyond LHY correlations are appreciable. 
Additionally, the droplet phase diagram in 1D optical lattices has been discussed within a density matrix renormalization group method explicating a transition from a Mott-insulator to a pair superfluid~\cite{morera2020quantum,morera2021universal}.

Interestingly, a major focus has been placed on the impact of intercomponent attraction when intracomponent repulsion is held fixed.
However, the interplay between the two, that provides an additional knob for droplet deformation remains unexplored at the many-body level. 
In the same spirit,  especially interesting
aspects of this competition can take place in 1D LHY fluids which occur
at the balance point where MF interactions are cancelled out~\cite{jorgensen2018dilute,skov2021observation,guo2021leehuangyang}.
Additionally, the properties of 1D droplets and their consequent deformations in mass-imbalanced mixtures are not yet studied. 
It is important to highlight here that much
of the current understanding regarding the dynamics of 1D droplets is presently
limited  to MGP predictions. 
Notable examples are droplet inelastic collisions~\cite{astrakharchik2018dynamics} and their merging into breathers~\cite{liu2019symmetry}, droplet nucleation triggered by modulational instability mechanism~\cite{mithun2020modulational}, 
among others; see, e.g.,~\cite{luo2021new} for a recent review. 
In view of the above, it is essential  to further study both the static properties and the nonequilibrium dynamics of droplets to reveal the necessity of beyond LHY corrections in order to adequately capture the inevitable build-up of interparticle many-body correlations. 

In the present work, we consider a particle balanced two-component 1D bosonic mixture with intracomponent repulsion and intercomponent attraction.
To address the ground state and the nonequilibrium quantum dynamics of the ensuing droplet states we use the {\it ab-initio} nonperturbative multi-layer multi-configuration time-dependent Hartree method for atomic mixtures (ML-MCTDHX)~\cite{cao2017unified,cao2013multi}. 
Within this approach, it is possible to encapsulate all the emergent intra- and intercomponent correlations of the mixture~\cite{mistakidis2018correlation,mistakidis2021radiofrequency,mistakidis2019quench}. 
To expose the role and degree of correlations 
on the droplet formation and dynamics we compare the variational results with the MF (absence of LHY corrections) and MGP (incorporating LHY corrections) models but also with a species mean-field (SMF) approach which naturally accounts for intracomponent correlations. 
Operating at distinct correlation levels we explain the observed alterations of  the droplet profiles, stemming from intercomponent correlations (entanglement) and showcase that they are more prominent for stronger intercomponent attraction or the presence of a harmonic trap.

At the ground state level, a transition from a spatially delocalized to a strongly localized Gaussian-shaped droplet is revealed upon increasing the intracomponent repulsion.
Additionally, the droplet's shape deforms from a Gaussian-like towards a FT structure for either fixed interactions and larger particle numbers~\cite{astrakharchik2018dynamics} or by solely considering smaller intercomponent attractions~\cite{parisi2020quantum}. 
Droplets exhibiting a FT experience a correlation hole which diminishes when they morph into Gaussian-like structures. 
For LHY fluids, a gradual reshaping to a narrower density distribution is realized for increasing atom numbers. 
Few-atom liquids are characterized by a correlation hole while for larger particle numbers a tendency for long-range two-body correlations is observed. 
Imposing a harmonic trap, FT signatures fade away. 
Again, the droplets transition from a delocalized to a localized configuration for stronger attraction. 
Similar structural droplet deformations take place for weakly mass-imbalanced mixtures~\cite{d2019observation,burchianti2020dual} where the components show an enhanced mixing, whilst for larger mass differences excitation patterns are identified. 

Next, we monitor for the first time the nonequilibrium droplet dynamics within an \textit{ab-initio} approach. 
Particularly, we exemplify the droplet expansion~\cite{rodriguez2021oscillating}
which deforms in the course of the evolution featuring 
FT signatures following quenches to stronger intercomponent attractions. 
In the reverse case, a breathing motion is initiated. 
The importance of the intercomponent attraction manifests itself in the slower expansion velocity and breathing frequency of the droplet, contrary to the MGP predictions.
The inclusion of a harmonic trap results in a breathing motion irrespective of the postquench value with a larger predicted frequency in the MGP approach. 
Therefore, both the expansion and the breathing dynamics of the droplet can be considered as sensitive experimental probes for exposing beyond LHY physics. 
Importantly, two-body correlations develop during the evolution with two bosons featuring an anti-correlation (antibunching) at the same location and a correlation tendency when  being further apart. 
Such correlation droplet patterns are yet to be 
monitored experimentally. 

This work is organized as follows. 
Section~\ref{sec:theory} presents the considered droplet setting and the various methodologies employed to describe the ensuing droplet configurations. 
The ground state correlation properties of droplets focusing on the impact of the participating interactions and particle number are analyzed for symmetric (mass-balanced) and asymmetric (mass-imbalanced) bosonic mixtures in Sec.~\ref{symmetric_drops} and \ref{assymetric_drops}, respectively. 
The emergent interaction quench dynamics of both symmetric and asymmetric mixtures after a sudden change of the intercomponent coupling is discussed in Sec.~\ref{quenches}. 
A summary of our findings and some relevant future perspectives 
 are provided in Section~\ref{sec:conclusions}.

\section{Droplet setting and theoretical approaches}\label{sec:theory} 
\subsection{The interacting Hamiltonian }\label{hamiltonian}

To realize droplet structures we consider a particle-balanced bosonic mixture with $N_A=N_B$ atoms residing in a box potential. 
Below, we mainly focus on the cases of both equal ($M_A=M_B$) and different ($M_A\approx 2.12 M_B$) masses which experimentally correspond to a homonuclear mixture of distinct $^{39}$K hyperfine states~\cite{cabrera2018quantum,semeghini2018self,cheiney2018bright} (e.g. $\ket{F=1, m_F=-1}$ and $\ket{F=1, m_F=0}$) and heteronuclear setups consisting of $^{41}$K and $^{87}$Rb (e.g. in the $\ket{F=1,m_F=1}$ state) isotopes~\cite{d2019observation,burchianti2020dual} respectively. 
The generic MB Hamiltonian of these mixtures reads 
\begin{equation}\label{eq:MB_Hamilt}
\begin{split}
&H = \sum_{\sigma = A, B}^{}\sum_{i = 1}^{N_\sigma} -\frac{\hbar^2}{2M_{\sigma}}\bigg(\frac{\partial}{\partial x_i^{\sigma}}\bigg)^2  
+ \sum_{\sigma=A,B}  g_{\sigma \sigma}\\&\sum_{ i \geq j }^{} \delta(x^{\sigma}_i - x^{\sigma}_j)    
+ g_{AB}\sum_{i=1}^{N_A} \sum_{j=1}^{N_{B}} \delta(x^{A}_i - x^{B}_j).
\end{split}
\end{equation}
Since we operate within the zero temperature limit, $s$-wave scattering is the dominant interaction process~\cite{olshanii1998atomic}.  
As such, the intra- and the intercomponent interactions are modeled by a contact potential characterized by the effective coupling constants $g_{AA}$, $g_{BB}$ and $g_{AB}$. 
The latter acquire the form ${g_{\sigma \sigma'}} =\frac{{2{\hbar ^2}{a^s_{\sigma \sigma'}}}}{{\mu a_ \bot ^2}}{\left( {1 - {\left|{\zeta (1/2)} \right|{a^s_{\sigma \sigma'}}}
/{{\sqrt 2 {a_ \bot }}}} \right)^{ -1}}$~\cite{olshanii1998atomic}. 
In this expression, $\mu=M_{\sigma}M_{\sigma'}/(M_{\sigma}+M_{\sigma'})$ is the reduced mass, $\zeta$ denotes the Riemann zeta function, $a_\perp$ is the transverse length scale, and ${a^s_{\sigma \sigma'}}$ refers to the three-dimensional $s$-wave intra- ($\sigma=\sigma'$) or intercomponent ($\sigma \neq \sigma'$) scattering length. 
Experimentally, $g_{\sigma\sigma'}$ is tunable either through ${a^s_{\sigma \sigma'}}$ utilizing Feshbach resonances~\cite{chin2010feshbach,kohler2006production} or by adjusting $a_{\perp}$ by means of confinement-induced resonances~\cite{olshanii1998atomic}. 

For the MB calculations, to be presented below, a box potential~\cite{gaunt2013bose} of finite length $L$ is considered by imposing hard-wall boundary conditions at $x_{\pm}=\pm 50$. 
The impact of finite size effects, whenever present, is discussed in the main text. 
In the following, we rescale our Hamiltonian in terms of $\hbar^2/(M_A L^2)$. 
The length, time and interactions are measured in terms of $L$, $(M_AL^2)/\hbar$ and $\hbar^2/(M_AL)$. 
Our 1D setting can be experimentally addressed e.g. when considering a $^{39}$K gas of $N=40$ bosons with $g_{AA}=g_{BB}= 0.1 \hbar^2/(M_AL)\approx  3.4\times 10^{-39}$ Jm in a box potential of length $L=5$~$\mu m$. 
Temperature effects are negligible for $ T \ll k_B^{-1} \hbar^2 N^2/(M_AL^2) \approx 1.5$ $\mu K$~\cite{pitaevskii2016bose}, with $k_B$ being the Boltzmann constant and $T$ is the temperature of the bosonic cloud. 
Also, typical evolution times of the order of $10^3$ correspond approximately to $\sim 0.6$~seconds. 

To elaborate on the correlation patterns of 1D droplet configurations, we study the ground state of the above-described bosonic mixture characterized by the same intracomponent repulsive interactions, i.e. $g_{AA}=g_{BB}>0$, and intercomponent attraction $g_{AB}<0$. 
In order to facilitate the discussion, we first define the parameters $g=\sqrt{g_{AA}g_{BB}}$ and 
$\delta g=g-\abs{g_{AB}}$ with the latter quantifying the balance point of MF attraction and intracomponent repulsion occurring when $\delta g=0$, see also Sec.~\ref{MGP_method}. 
A basic condition for the formation of quantum droplets is $\abs{\delta g}/g\ll 1$~\cite{petrov2016ultradilute,petrov2015quantum}. Concretely, we shall unravel the impact of the involved interactions by varying $g$ while keeping fixed the balance point $\delta g/g$ and vice versa, i.e. tuning $\delta g/g$ with $g$ constant. 
This will allow us to realize a transition from Gaussian to FT droplet structures. Notice that since a mesoscopic number of atoms is considered, it is not possible to create pronounced FT profiles.  Interestingly, two different settings will be analyzed namely the SU(2) symmetric mass-balanced ($M_A=M_B$) bosonic mixture [Sec.~\ref{symmetric_drops}] and the asymmetric mass-imbalanced ($M_A\neq M_B$) one [Sec.~\ref{assymetric_drops}]. 
It can be readily deduced that in the former case the two components behave identically and therefore it is sufficient to examine only one of them, while in the latter scenario the mass-imbalance renders the involved species distinguishable exhibiting a different configuration and hence 
requiring a genuinely two-component description. 

After investigating the emergent structural shape of the droplets, we shall proceed to study their interaction quench dynamics following sudden changes of the control parameter $\delta g/g$ with $g$ fixed and thus adjusting solely $g_{AB}$. 
More specifically, for increasing $\delta g/g$ we will probe the dynamical transition from a Gaussian to a FT profile while the droplet 
expands. 
For the reverse quench scenario a collective breathing motion of the droplet is excited. 
In all cases, careful comparisons between a variational and different mean-field type approaches [see Sec.~\ref{variational_method}, \ref{MGP_method}] including quantum fluctuations at different levels, reveal the crucial role of the underlying correlations and in particular the intercomponent ones.

\subsection{Variational method and wave function reductions}\label{variational_method} 

To address the ground state correlation properties and most importantly the interaction quench dynamics of the droplet setting described by Eq.~(\ref{eq:MB_Hamilt}) we solve the time-dependent MB Schr{\"o}dinger equation. For this we mainly rely on the variational ML-MCTDHX method~\cite{cao2017unified,cao2013multi}.  
A key facet of this approach is that the total MB wave function of the underlying multicomponent system is expanded in terms of a time-dependent and variationally optimized basis set~\cite{lode2020colloquium}. 
The latter enables us to capture all the emergent intra- and intercomponent correlations~\cite{mistakidis2018correlation,katsimiga2017dark} as well as adjusting our MB ansatz to operate at specific correlation orders, a procedure that we explicate below.  

In particular, the intercomponent correlations (entanglement) are taken into account by expressing the MB wave function, $|\Psi(t)\rangle$, according to a truncated Schmidt decomposition~\cite{horodecki2009quantum} consisting of $D$ different species functions, $|\Psi^{\sigma}_k(t)\rangle$, for each component $\sigma=A,B$, as follows   
\begin{equation} 
|\Psi(t)\rangle=\sum_{k=1}^D \sqrt{\lambda_k(t)} |\Psi^{\rm A}_k(t)\rangle|\Psi^{\rm B}_k(t)\rangle.  
\label{eq:wfn_total}
\end{equation} 
The time-dependent expansion coefficients $\lambda_k(t)$ are the corresponding Schmidt weights and are also known as the natural populations of the $k$-th species function.  
Subsequently, the system is referred to as intercomponent correlated (entangled)~\cite{horodecki2009quantum,roncaglia2014bipartite} if at least two distinct $\lambda_k(t)$ exhibit a nonzero population; otherwise when $\lambda_1(t)=1$ and $\lambda_{k>1}(t)=0$, it is non-entangled. 
In the latter case intercomponent entanglement is ignored and the total wave function acquires the following tensor product form of two states   
\begin{equation} 
|\Psi_{\rm{SMF}}(t)\rangle=|\Psi^{\rm A}_1(t)\rangle \otimes |\Psi^{\rm B}_1(t)\rangle.  
\label{eq:wfn_SMF}
\end{equation} 
This situation constitutes a reduction of the MB ansatz being commonly referred to as the species mean-field (SMF) approximation~\cite{mistakidis2019effective,cao2017unified}. 
Within this framework only the intracomponent correlations are considered by further expanding the individual species wave functions with respect to a time-dependent basis set. 

Next, in order to incorporate intracomponent correlations, each species function appearing either in Eq.~(\ref{eq:wfn_total}) or Eq.~(\ref{eq:wfn_SMF}) is written as a linear superposition of time-dependent number-states, $|\vec{n} (t) \rangle^{\sigma}$, namely 
\begin{equation}
| \Psi_k^{\sigma} (t) \rangle =\sum_{\vec{n}} C^{\sigma}_{k;\vec{n}}(t) | \vec{n} (t) \rangle^{\sigma}.  
\label{eq:wfn_numb_states}
\end{equation} 
Here, $C^{\sigma}_{k;\vec{n}}(t)$ denote the time-dependent expansion coefficients. 
The number state $|\vec{n} (t) \rangle^{\sigma}$ corresponds to a permanent building upon $d_{\sigma}$ time-dependent variationally optimized single-particle functions (SPFs) $\left|\phi_i^{\sigma} (t) \right\rangle$, with $i=1,2,\dots,d_{\sigma}$ and occupation numbers $\vec{n}=(n_1,\dots,n_{d_{\sigma}})$. Finally, these SPFs are expanded on a time-independent primitive basis which in our case corresponds to an $\mathcal{M}$ dimensional discrete variable representation with $\mathcal{M}=1500$ grid points. 
Another interesting feature of the ML-MCTDHX ansatz is that it can be naturally reduced to the usual MF one when all correlations of the multicomponent setup are absent~\cite{pitaevskii2016bose}. This is accomplished by assuming a single Schmidt coefficient and one SPF per species, thereby $D=d_A=d_B=1$. Then, all particles of a particular species solely occupy a single wave function 
\begin{equation} 
\ket{\Psi_{\rm {MF}}(t)}=
\prod_{i=1}^{N_A} \ket{\phi_i^{A} (t)}\prod_{i=1}^{N_B} \ket{\phi_i^{B} (t)}.  
\label{eq:wfn_MF}
\end{equation}
It yields a coupled set of Gross-Pitaevskii equations for the bosonic mixture~\cite{kevrekidis2007emergent,pitaevskii2016bose}. A comparison of the predictions of this approach to the MB one will shed light onto the impact of interparticle correlations on the droplet formation.

Having at hand, the desired form of the wave function ansatz as described above by Eqs.~(\ref{eq:wfn_total}), (\ref{eq:wfn_SMF}), (\ref{eq:wfn_numb_states}) we determine the respective ML-MCTDHX equations of motion~\cite{cao2017unified} aiming to numerically solve the underlying Schr{\"o}dinger equation and thus calulate the ($N_A+N_B$)-body wave function satisfying the Hamiltonian of Eq.~(\ref{eq:MB_Hamilt}).
This is achieved by following a variational principle, e.g. the Dirac-Frenkel one \cite{frenkel1934wave,dirac_1930} for the above introduced generalized ansatz. 
It leads to a set of $D^2$ linear differential equations of motion for the $\lambda_k(t)$ coefficients being coupled to $D\big(\frac{(N_A+d_A-1)!}{N_A!(d_A-1)!}+\frac{(N_B+d_B-1)!}{N_B!(d_B-1)!}\big)$ nonlinear integrodifferential equations for the species functions and $d_A+d_B$ nonlinear integrodifferential equations for the SPFs. For instance, applying the Dirac-Frenkel variational priniciple to Eq.~(\ref{eq:wfn_SMF}) the resulting equations of motion within the SMF method read 
\begin{equation}
\begin{split}
&i\hbar \frac{\partial \Psi_A(x)}{\partial t}-\Lambda_B(t)\Psi_A(x)=H_A\Psi_A(x)+g_{AB}\rho_B^{(1)}(x) \Psi_A(x),\\  
&i\hbar \frac{\partial \Psi_B(x)}{\partial t}-\Lambda_A(t)\Psi_B(x)=H_B\Psi_B(x)+g_{AB}\rho_A^{(1)}(x) \Psi_B(x).  
\end{split}\label{SMF_equations}
\end{equation}
Here, $\Lambda_{\sigma}(t)=\int dx \Psi^*_{\sigma}(x)[H_{\sigma}\Psi_{\sigma}(x)-i\hbar\partial_t \Psi_{\sigma}(x)]$ with $H_{\sigma}=-(\hbar^2/(2M_{\sigma}))\sum_{i=1}^{N_{\sigma}}\partial^2_{x_i} +g_{\sigma \sigma}\sum_{i<j} \delta(x_i-x_j)$ and $\Psi_{\sigma}(x;t)\equiv \Psi_{\sigma}(x)$. 
Further details regarding the SMF wave function and equations are discussed in Refs.~\cite{cao2017unified,mistakidis2019effective}.

Concluding, the time-dependence of the used basis states allows us to efficiently truncate the Hilbert space of the system while accounting for all relevant intra- and intercomponent correlations utilizing a computationally feasible basis size. 
This way, the number of the involved equations of motion that are numerically solved remains tractable even for mesoscopic particle numbers. 
The Hilbert space truncation is designated by the employed orbital configuration space $C=(D;d_A;d_B)$ being for the systems to be considered below $C=(15;6;6)$ ($C=(15;5;5)$) for $N_A=N_B<20$ ($N_A=N_B>20$). 

\begin{figure*}[ht]
\includegraphics[width=0.9\textwidth]{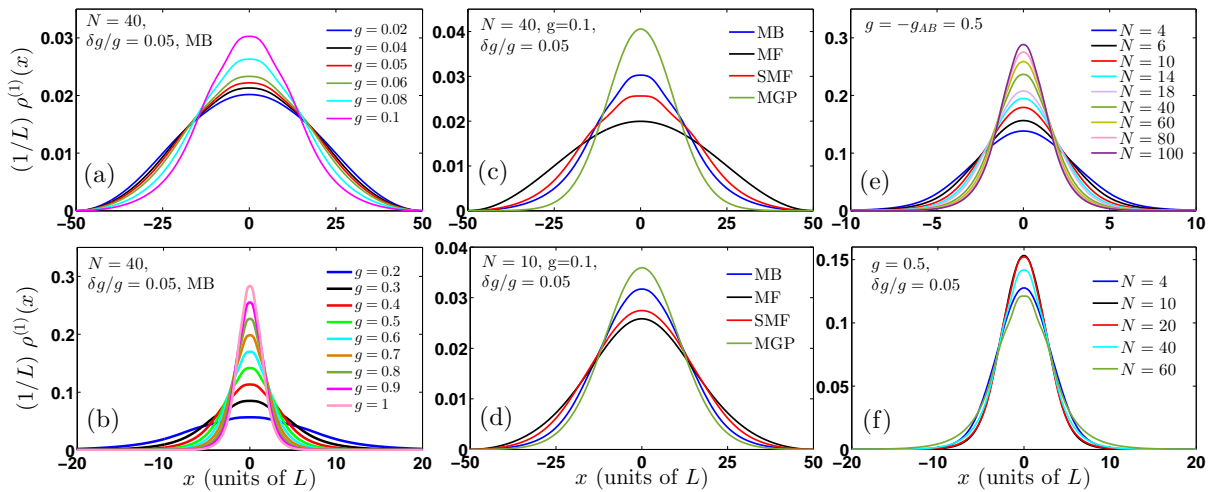}
\caption{Ground state densities $\rho^{(1)}(x)$ of a symmetric bosonic mixture for varying intracomponent interaction strength $g$ and particle number $N$ (see legends) representing various droplet configurations. (a), (b) $\rho^{(1)}(x)$ for different $g$ while keeping $\delta g/g =0.05$ fixed as obtained in the MB approach. The droplet profile features a gradual shrinking for larger $g$ and shows FT signatures around $g=0.1$.  Comparison of $\rho^{(1)}(x)$ for (c) $N=40$ and (d) $N=10$ between different approaches (see legend) explicating the impact of beyond LHY correlations as imprinted in the density of the droplet. 
$\rho^{(1)}(x)$ for increasing particle number $N$ of (e) an LHY fluid ($\delta g=0$) and (f) a droplet for fixed $g=0.5$ and $\delta g/g =0.05$. 
The fluid experiences a gradual localization for increasing $N$. 
The system consists of $N_A=N_B\equiv N/2$ bosons of equal mass and intracomponent interactions $g_{AA}=g_{BB} \equiv g$ trapped in a box potential of length $L$. }
\label{fig:den_sym_varyg} 
\end{figure*}

\subsection{Modified Gross-Pitaevskii approach}\label{MGP_method} 

A commonly used framework to describe the stationary and dynamical aspects of quantum droplets is the so-called MGP method~\cite{petrov2015quantum,petrov2016ultradilute}. 
It includes the effect of quantum fluctuations to leading order within the local density approximation. Indeed, it has been argued~\cite{mithun2020modulational,mithun2021statistical} that 1D bosonic droplet configurations in particle balanced mixtures with different intracomponent interactions $g_{AA}\neq g_{BB}$ but equal masses ($M_A=M_B\equiv M$) are described by the following set of coupled MGP equations of motion    
\begin{equation}
\begin{split}
&i \hbar \frac{\partial \Psi_A}{\partial t}= - \frac{\hbar^2}{2M} \frac{\partial^2 \Psi_A}{\partial x^2}+ (g_{AA}+G g_{BB}) \abs{\Psi_A}^2 \Psi_A\\&-(1-G)g\abs{\Psi_B}^2 \Psi_A- \frac{g_{AA} \sqrt{M}}{\pi \hbar} \\& \times \sqrt{g_{AA} \abs{\Psi_A}^2+g_{BB}\abs{\Psi_B}^2} \Psi_A,\\ 
&i \hbar \frac{\partial \Psi_B}{\partial t}= - \frac{\hbar^2}{2M} \frac{\partial^2 \Psi_B}{\partial x^2}+ (g_{BB}+G g_{AA}) \abs{\Psi_B}^2 \Psi_B\\&-(1-G)g\abs{\Psi_A}^2 \Psi_B- \frac{g_{BB} \sqrt{M}}{\pi \hbar} \\& \times \sqrt{g_{AA} \abs{\Psi_A}^2+g_{BB}\abs{\Psi_B}^2} \Psi_B.\label{eq:MGP_asym}
\end{split}
\end{equation} 
Here, $G=2g \delta g/(g_{AA}+g_{BB})^2$ quantifies deviations with respect to the threshold of MF repulsion and attraction which corresponds to $\delta g=g-\abs{g_{AB}}=0$ where $g=\sqrt{g_{AA}g_{BB}}$. 
Importantly, the creation of quantum droplets is ensured for repulsive intracomponent interactions ($g_{AA}>0$, $g_{BB}>0$) and attractive intercomponent ones ($g_{AB}<0$) which satisfy the condition $\delta g/ g \ll1$~\cite{petrov2016ultradilute}. 
Notice also that $\delta g/g$ is experimentally tunable with the aid of Feshbach resonances~\cite{cabrera2018quantum,cheiney2018bright}, thus further motivating the relevance of considering quenches of this parameter to induce nonequilibrium droplet dynamics. 

It is also worth mentioning that in the case of a perfectly symmetric mixture, assuming $M_A=M_B$, $g_{AA}=g_{BB}$ and $N_A=N_B$, the above coupled set of MGP Eqs.~(\ref{eq:MGP_asym}) reduces to a single-component equation~\cite{petrov2016ultradilute,astrakharchik2018dynamics}. 
The latter depending on the sign of $\delta g/g$ assumes different solutions. 
Particularly, for $\delta g/g>0$ there is a localized Gaussian-shaped solution for small particle numbers turning into a FT profile for increasing $N$. 
These configurations represent a quantum droplet generated by the balance of (dispersion with) the effective cubic self-repulsion and quadratic attraction. This is the regime that our analysis is based on. 
We remark, however, that if $\delta g /g<0$ there are various solitonic solutions, see for a more detailed discussion Ref.~\cite{mithun2020modulational}. 
Concluding, let us emphasize that a particular focus of our study will be the comparison of the above-mentioned approaches, namely the \textit{ab-initio} one [Eq.~(\ref{eq:wfn_total}), (\ref{eq:wfn_numb_states})] the SMF [Eq.~(\ref{SMF_equations})], the common MF [Eq.~(\ref{eq:wfn_MF})] and the MGP [Eq.~(\ref{eq:MGP_asym})]. 

Another important point is that the MGP is known to provide an adequate description of droplets in the limit of large systems associated with high particle densities, smooth spatial variation and in the weakly interacting case. 
The latter is quantified via the Lieb-Liniger parameter $\gamma=mg/(\hbar^2 \rho_0) \ll 1$ where $\rho_0$ is the peak density of the gas. 
In our case where we operate with mesoscopic atom numbers typically $N_A=N_B=20$ and $N_A=N_B=50$ while $g=0.1$ we have $\gamma \approx 0.25$ and $\gamma \approx 0.1$ respectively. Also, owing to the considered relatively low particle number the gas density is not smoothly varying in space. 
Therefore, the used parameter regime lies beyond the validity of the MGP approach. However, as we shall argue below, even in this case the MGP provides valuable insights on the stationary and dynamical properties of mesoscopic droplets. For increasing atom numbers, and as 
$\rho_0$ increases/$\gamma$ decreases, the MGP description 
becomes a progressively more adequate one (yet this is outside
the regime that we can monitor with the MB computation herein). 
\begin{figure*}[ht]
\includegraphics[width=0.8\textwidth]{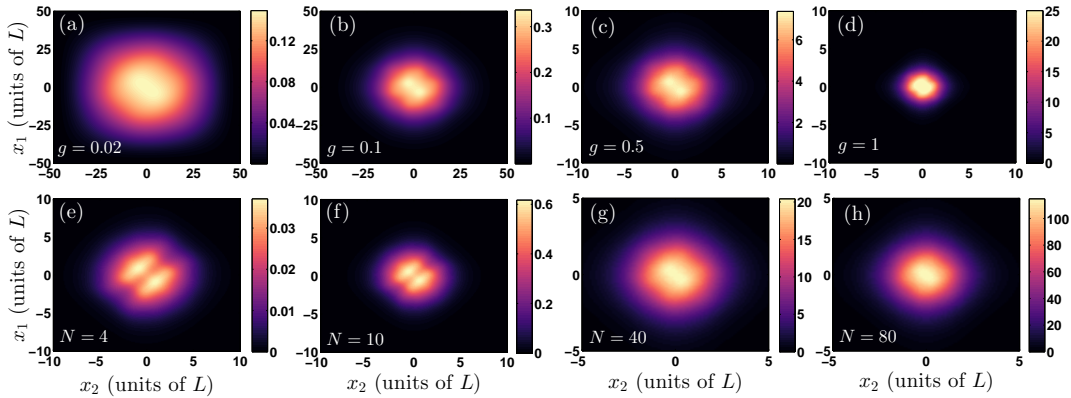}
\caption{Two-body density distributions of a symmetric bosonic mixture within the MB approach for different (a)-(d) intracomponent interactions $g$ while $\delta g/g=0.05$ and $N=40$. The distributions become spatially localized for increasing $g$ and a correlation hole emerges [(b), (c) panels] at interaction regimes where FT signatures appear in the density profile, see Fig.~\ref{fig:den_sym_varyg}(a), (b). (e)-(h) The same as in (a)-(d) but for a varying particle number $N$ of an LHY fluid with $g=0.5$ and $\delta g=0$. 
A two-body correlation hole is evident fading away for $N>10$.}
\label{fig:2bcor_sym_varyg} 
\end{figure*}

\section{Symmetric droplet configurations}\label{symmetric_drops}

We now discuss the role of MB effects on the formation of 1D droplet structures in the ground state of symmetric bosonic mixtures, namely when the components are characterized by the same mass ($M_A=M_B\equiv M$), particle number ($N_A=N_B$) and repulsive intracomponent interactions ($g_{AA}=g_{BB}>0$). 
The development of droplets is achieved by tuning the intercomponent coupling to attractive values and especially close to the balance point of MF attraction and repulsion being $\delta g=\sqrt{g_{AA}g_{BB}}-\abs{g_{AB}}=0$~\cite{petrov2016ultradilute,petrov2015quantum}. 
A main objective of our investigation is to elucidate the interplay of the intra- and intercomponent interactions on the structural shape and correlation patterns of the emergent droplet configurations probing also the transition from Gaussian-like distributions to a FT. 
Additionally, we shall expose the role of correlations at different orders for the adequate description of droplets by directly comparing the predictions of different approximations, namely the MF and SMF methods, the MGP and the MB approaches. 
To shed light into the crossover from few- to many-atom bound states we study bosonic mixtures close ($\delta g/g\ll1$) and at the threshold ($\delta g=0$) of the MF attraction and repulsion giving rise to droplet phases and LHY fluids respectively~\cite{skov2021observation}. 

To characterize the stationary and dynamical spatially resolved features of the droplet configurations we resort to the single-particle reduced density matrix of each $\sigma$-component~\cite{sakmann2008reduced,mistakidis2018correlation} 
\begin{equation}
\rho_\sigma^{(1)}(x,x';t)=\langle\Psi(t)|\hat{\Psi}_{\sigma}^{\dagger}(x)\hat{\Psi}_{\sigma}(x')|\Psi(t)\rangle.\label{eq:SPDM}
\end{equation}
The bosonic field operator $\hat{\Psi}_{\sigma}(x)$ of the $\sigma$-component acts at position $x$~\cite{pitaevskii2016bose}. 
In the following, we will in particular monitor the diagonal part of this observable, namely $\rho_\sigma^{(1)}(x;t)\equiv\rho_\sigma^{(1)}(x,x'=x;t)$, known as the one-body density of the $\sigma$-component which we consider as normalized to unity. 
It is experimentally tractable through an average over a sample of single-shot images~\cite{bloch2008many,mistakidis2018correlation}. Notice that for symmetric mixtures $\rho_A^{(1)}(x;t)=\rho_B^{(1)}(x;t)\equiv \rho^{(1)}(x;t)$.

\subsection{Interplay of intra- and intercomponent interactions}\label{vary_g}

As a first step we investigate the impact of the intra- and intercomponent interaction coefficients on the structural shape of $\rho^{(1)}(x)$. To this end, focusing on a particular particle number e.g. $N=40$ we inspect $\rho^{(1)}(x)$ for increasing $g$ while keeping $\delta g/ g$ fixed, see Fig.~\ref{fig:den_sym_varyg} (a), (b). Evidently, $\rho^{(1)}(x)$ features a delocalized shape for adequately weak $g$, it possesses FT signatures for intermediate couplings (e.g. $g=0.1$) and finally for larger values of $g>0.5$ it shrinks further forming a localized Gaussian. This behavior is attributed to the fact that a constant $\delta g/g$ but increasing $g$ is associated to larger intercomponent attractions which are responsible for this gradual reshaping of $\rho^{(1)}(x)$ from a more delocalized to a 
progressively localized Gaussian distribution. 
Note that a similar transition behavior in $\rho^{(1)}(x)$ occurs as well for other particle numbers, e.g., $N=10,20$. 
However, the signatures of the FT are suppressed especially for $N<10$ (not shown). 
\begin{figure*}[ht]
\includegraphics[width=0.8\textwidth]{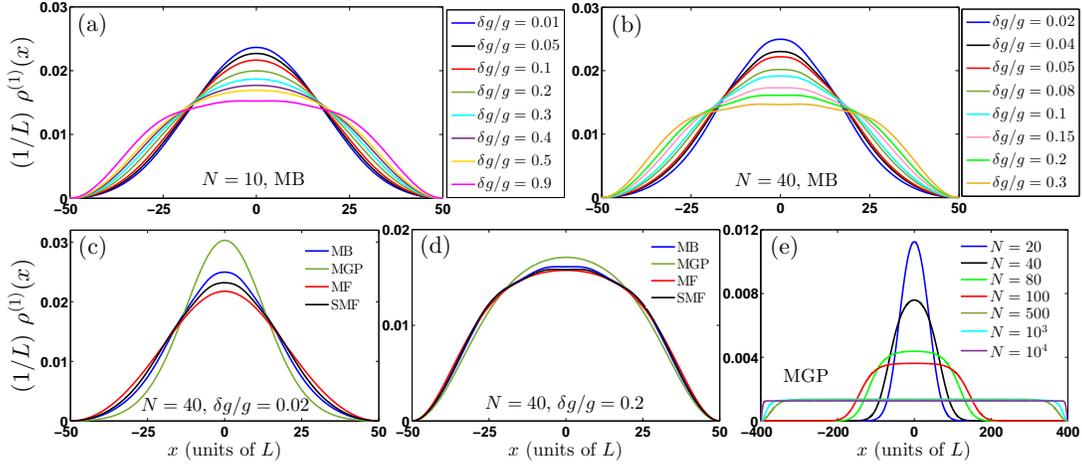}
\caption{Ground state densities $\rho^{(1)}(x)$ of a symmetric bosonic mixture for fixed $g=0.05$ and varying $\delta g /g$ (see legend) for (a) $N=10$ and (b) $N=40$ within the MB approach. 
A transition from a Gaussian-shaped to a FT profile occurs for increasing $\delta g/g$. This transition is realized for smaller $\delta g/g$ values when $N$ is larger. 
Comparison of $\rho^{(1)}(x)$ for $N=40$ between different methods (see legend) for $g=0.05$ while (c) $\delta g /g=0.02$, and (d) $\delta g /g=0.2$. 
It becomes clear that beyond LHY correlations are suppressed for larger $\delta g/g$ since then $g_{AB} \to 0$. 
(e) $\rho^{(1)}(x)$ as predicted by the MGP for varying $N$ with $g=0.05$ and $\delta g/g=0.2$. The droplet profile becomes delocalized for increasing $N$ acquiring a FT profile which saturates for $N>500$. 
The symmetric bosonic mixture comprises of $N_A=N_B\equiv N/2$ bosons with the same mass and intracomponent interactions $g_{AA}=g_{BB} \equiv g$ while it experiences a box potential of length $L$.}
\label{fig:den_sym_varydg} 
\end{figure*} 

This deformation of $\rho^{(1)}(x)$ can also be observed within the MGP framework but showing qualitative deviations from the MB case (see below). 
In the MGP limit the modified healing length~\cite{tylutki2020collective} of the mixture reads 
\begin{equation}
 \xi=\frac{\pi\hbar^2}{M}\frac{\sqrt{\abs{\delta g}}}{g\sqrt{2g}}. \label{mod_healing}
\end{equation}
According to this expression, when $\delta g/g$ is held constant and $g$ is varied to larger values then $\xi$ decreases which explains the above-described deformation of $\rho^{(1)}(x)$. 
Notably, within the MGP the above phenomenology ceases to exist for $N>800$ and $\rho^{(1)}(x)$ acquires a pronounced FT profile as $g$ increases. 
This suggests the involvement of finite-size effects, to be investigated in future studies. 
We also remark that the above-discussed FT signatures of $\rho^{(1)}(x)$ at intermediate interactions are not present within MGP. 
Therefore, they constitute an intriguing prospect to identify beyond LHY physics for mesoscopic particle numbers.

To track the related two-body spatially resolved correlation patterns characterizing the ground state of the droplet structures we determine the diagonal of the intracomponent two-body reduced density matrix
\begin{equation}
\rho^{(2)}_{\sigma \sigma}(x_1,x_2;t)=\langle\Psi(t)|\hat{\Psi}_{\sigma}^{\dagger}(x_2)\hat{\Psi}^{\dagger}_{\sigma}(x_1)\hat{\Psi}_{\sigma}(x_1)\hat{\Psi}_{\sigma}(x_2)|\Psi(t)\rangle. \label{2body_reduced}
\end{equation} 
This represents the probability to simultaneously detect two $\sigma$-component bosons at positions $x_1$ and $x_2$ respectively and for symmetric mixtures $\rho^{(2)}_{AA}(x_1,x_2;t)=\rho^{(2)}_{BB}(x_1,x_2;t) \equiv \rho^{(2)}(x_1,x_2;t)$. 
A variety of two-body configurations is presented in Figs.~\ref{fig:2bcor_sym_varyg} (a)-(d) for specific values of $g$, see also Fig.~\ref{fig:den_sym_varyg} (a), (b). 
It becomes apparent that at the two-body level the aforementioned transition is characterized by a delocalized circularly symmetric two-body density distribution, e.g., for weak $g=0.02$ [Fig.~\ref{fig:2bcor_sym_varyg} (a)] which gradually shrinks in space for a larger $g$, see Figs.~\ref{fig:2bcor_sym_varyg} (b)-(d). 
Notice also that for interactions where $\rho^{(1)}(x)$ exhibits signatures of a FT profile we observe a tendency for the formation of a correlation hole~\cite{mukherjee2020induced,keiler2020doping} along the diagonal of $\rho^{(2)}(x_1,x_2)$ [Figs.~\ref{fig:2bcor_sym_varyg} (b), (c)]. This implies that two bosons avoid to reside together in the region of the FT, i.e. the latter experiences an anti-correlated behavior. 
\begin{figure*}[ht]
\includegraphics[width=0.9\textwidth]{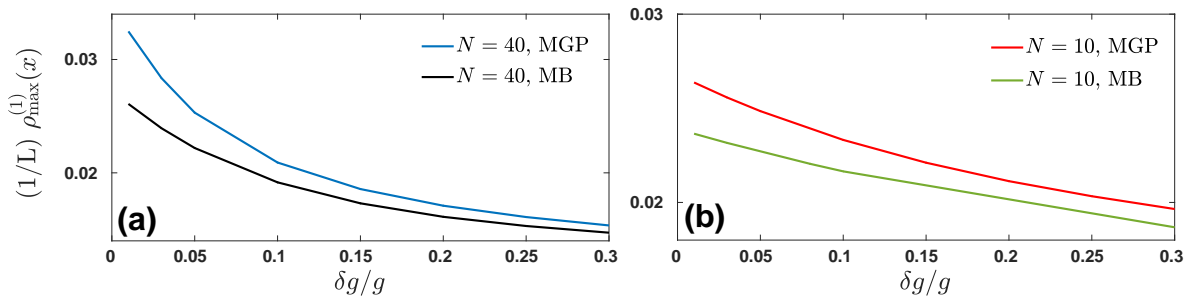}
\caption{Density maximum of the ground state droplet with respect to $\delta g/g$ within the MB and the MGP frameworks as well as for different particle numbers, i.e. (a) $N=40$ and (b) $N=10$. Deviations caused by beyond LHY correlations range from $\sim25\%$ ($\sim11\%$) to $\sim 4\%$ ($\sim5\%$) for $N=40$ ($N=10$). 
The mass-balanced bosonic mixture has intracomponent interactions $g_{AA}=g_{BB}\equiv g=0.05$.}
\label{fig:peak_den_comparison} 
\end{figure*} 

To understand the role of correlations in the formation of the ensuing droplet structures we contrast the predictions of different approaches, namely the MB, the MGP, the SMF and the MF methods. 
They operate at different correlation orders as explained in Sec.~\ref{variational_method}, \ref{MGP_method}. 
As a characteristic example, we concentrate on the case of $g=0.1$ and $\delta g/g =0.05$ for $N=40$ [Fig.~\ref{fig:den_sym_varyg} (c)] and $N=10$ [Fig.~\ref{fig:den_sym_varyg} (d)].  Overall, the MB result differs significantly from the one obtained within the aforementioned MF-type approaches~\footnote{The mean-field interaction parameter $Ng$ increases for larger $N$ while keeping $g$ fixed, thus rendering the gas stronger interacting. This explains the more prominent deviation between the MB and the MF approaches, compare Figs.~\ref{fig:den_sym_varyg} (c), (d).}.
It can be readily seen that in both cases, the MF outcome significantly deviates from 
the actual density configuration of the droplet showing a delocalization when compared to the MB case. 
This evinces the necessity of quantum fluctuations in this interaction regime. 
Recall, that in 1D quantum fluctuations are attractive~\cite{petrov2016ultradilute}. 
Interestingly, while capturing the overall shape of the droplet configuration, the MGP always overestimates its amplitude exhibiting accordingly a smaller width. 
This deviation explicates the presence of residual higher-order correlation effects, an observation that has also been exemplified in terms of quantum Monte-Carlo~\cite{parisi2020quantum}. 
Recall that due to the considered mesoscopic atom number the MGP prediction is not expected to be {\it a-priori} valid. 
On the other hand, the SMF method predicts a wider distribution than the MB implying that intercomponent entanglement plays indeed a crucial role. 
The latter becomes more prominent for larger values of $g$ and $\delta g/g$ constant rendering the MGP predictions less accurate. 

Moreover, we study the behavior of $\rho^{(1)}(x)$ at the balance point of MF attraction and repulsion, i.e. $\delta g=0$, for increasing particle number [Fig.~\ref{fig:den_sym_varyg} (e)]. 
Recall that in this limit the MF interactions cancel and the bosonic mixture is governed by the LHY correction term, see also Eq.~(\ref{eq:MGP_asym}). 
This way, we are able to explicitly visualize the crossover from few- to many-atoms at the threshold where the so-called LHY fluid forms~\cite{jorgensen2018dilute}. 
The latter bears a recent experimental realization~\cite{skov2021observation} for the hyperfine states $\ket{F=1,m_F=-1}$ and $\ket{F=1,m_F=0}$ of $^{39}$K offering available Feshbach resonances for approaching the threshold $\delta g=0$~\cite{cabrera2018quantum,semeghini2018self}. 
It can be deduced that the density of the mixture becomes narrower for increasing $N$ and in particular $\rho^{(1)}(x)$ tends to a more localized Gaussian distribution. 
A similar crossover takes place also within the MGP approach,  at least up to $N\sim 10^6$ atoms that we have checked, but having a relatively smaller width of the density profile.  
This is in sharp contrast to the behavior exhibited for $\delta g/g \ll 1$ where an increasing particle number with $g$ and $\delta g/g$ constant leads to a deformation from a Gaussian-like configuration first to a narrower one and then to a wider profile showing signatures of a FT for $N>50$, see Fig.~\ref{fig:den_sym_varyg}(f). 
The occurrence of such a FT in this regime of $N$ has also been reported using a quantum Monte-Carlo approach in Ref.~\cite{parisi2020quantum}.
This phenomenon is equally captured by the MGP framework where a pronounced FT builds upon $\rho^{(1)}(x)$ for $N>400$ within the parameter region that we operate, a particle number that is not easily tractable in the MB approach. 

Inspecting the corresponding two-body reduced density configurations of the LHY fluids for small particle numbers [Figs.~\ref{fig:2bcor_sym_varyg} (e), (f)] we observe the formation of a correlation hole across the diagonal and a two-hump structure along the anti-diagonal. 
This implies that two bosons avoid to reside in the same position but rather prefer to lie symmetrically with respect to $x=0$. Interestingly, the correlation hole is suppressed for larger particle numbers [Figs.~\ref{fig:2bcor_sym_varyg} (g), (h)] meaning that there is a finite probability for two atoms to approach one another around $x=0$. Additionally, a noticeable elongation of $\rho^{(2)}(x_1,x_2)$ with respect to $x_1=0$ and $x_2=0$ takes place which suggests the presence of long-range two-body correlations in the system.

\subsection{Tunability in terms of intercomponent attraction and finite size}\label{vary_dg}

Subsequently, we examine the structural deformations of droplets due to deviations from the MF balance point ($\delta g=0$) but retaining the condition $\delta g/g >0$. 
In this sense we fix $g=0.05$ and vary the intercomponent coupling $g_{AB}$ or equivalently $\delta g/g$. Notice that in this context an increasing $\delta g/g$ parameter corresponds to a decreasing magnitude of $g_{AB}$. Characteristic density profiles of the mixture in the MB approach are presented in Fig.~\ref{fig:den_sym_varydg} (a) and (b) for small ($N=10$) and larger ($N=40$) particle numbers respectively. In both cases, a transition from a highly delocalized Gaussian-shaped (for weak $\delta g/g$) to a more prominent FT density profile~\cite{astrakharchik2018dynamics,parisi2020quantum} (for larger $\delta g /g$) takes place. The relevant transition occurs for larger values of $\delta g/g$ when few atomic ensembles (e.g. $N=10$) are considered and in this latter case the FT structure is arguably less pronounced. 
Notice that for $N<10$ the FT configuration is basically absent. 
Actually, it can be shown that for increasing $\delta g/g$ and constant $g$ the FT configuration experiences a saturation tendency. 
This behavior is explained within the MGP framework since the corresponding energy functional exhibits a minimum~\cite{parisi2020quantum} at a density  
\begin{equation}
n_{0}=\frac{M}{N \pi^2 \hbar^2} \frac{[(2g-\delta g)^{3/2}+(\delta g)^{3/2}]^2}{(\delta g)^2}.\label{equil_den}
\end{equation}
It can be readily seen that in the limit of $\delta g \to g$ it holds that $n_{0}=4Mg/(N\pi^2 \hbar^2)$. We remark that close to the aforementioned transition boundary finite size effects come into play in our system; otherwise they are negligible. 
This means that by employing periodic boundary conditions or a larger length of the box potential the FT shape of the corresponding configurations becomes more pronounced. 
\begin{figure*}[ht]
\includegraphics[width=0.9\textwidth]{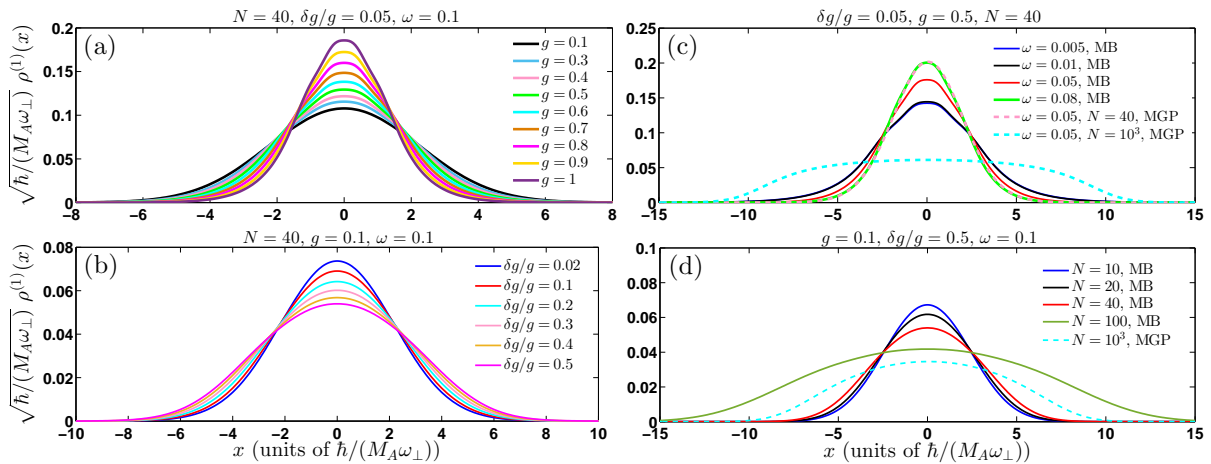}
\caption{Ground state densities $\rho^{(1)}(x)$ of a harmonically trapped bosonic mixture with $N=40$ for different (a) $g$ with fixed $\delta g/ g=0.05$ and (b) $\delta g /g$  with constant $g=0.1$ in the MB approach. The trap leads to localized density distributions experiencing a decreasing width for larger (smaller) values of $g$ ($\delta g/g$). (c) $\rho^{(1)}(x)$ for various frequencies $\omega$ of the harmonic trap (see legend) while $g=0.5$ and $\delta g /g=0.05$ are kept fixed. 
Evidently, a tighter trap results in the spatial localization of the droplet density distribution. 
(d) $\rho^{(1)}(x)$ for different particle numbers when $g=0.1$ and $\delta g /g=0.5$. 
The density profiles become wider for larger atom numbers. 
The dashed lines in (c), (d) showcase specific MGP predictions which exemplify significant deviations from the MB case. 
The symmetric mixture comprises of $N_A=N_B\equiv N/2$ bosons of the same mass and intracomponent interactions $g_{AA}=g_{BB} \equiv g$ in a harmonic trap of frequency $\omega$.}
\label{fig:trap_den_sym} 
\end{figure*} 

To appreciate the role of correlations in the formation of the above-described structures we explore $\rho^{(1)}(x)$ in the case of $N=40$ for weak $\delta g/g=0.02$ [Fig.~\ref{fig:den_sym_varydg} (c)] and strong $\delta g/g=0.2$ [Fig.~\ref{fig:den_sym_varydg} (d)] within different approaches. Focusing on $\delta g/g=0.02$ we observe that the MGP predicts a prominently more localized density distribution than the MB scenario, see also Fig.~\ref{fig:peak_den_comparison} (a). Interestingly, both the MF and the SMF methods capture better (in comparison
to the MGP case) the overall shape of $\rho^{(1)}(x)$ as can be seen in Figs.~\ref{fig:den_sym_varydg} (c), (d). However, they underestimate the value of the respective density maximum evincing that intercomponent entanglement plays indeed a role. Turning to $\delta g/g=0.2$ it can be easily seen that the predictions of the MF and the SMF methods are indeed in good agreement with the MB ones while the MGP result shows a slightly larger density peak. 
The fact that the droplet peak density is lower in the MB than in the MGP case holds independently of the particle number while the relevant deviation diminishes for increasing $\delta g/g$ with $g$ fixed since $g_{AB}$ becomes less attractive, see Fig.~\ref{fig:peak_den_comparison}. 
The observation that the deviations of the SMF approach to the MB one are reduced for a larger $\delta g /g$ is corroborated by the fact that in this case a weaker, and in particular close to zero, intercomponent interaction is involved and thus entanglement vanishes. Apparently, in this case also intracomponent correlations are weak since the MF prediction lies close to the MB. 

Additionally, we should mention that for $\delta g/g<0.4$ the density peak of the MGP approach is larger from the one obtained in the MB framework, while this situation is reversed for $\delta g/g>0.4$ (not shown). This behavior has been observed also within the realm of quantum Monte-Carlo~\cite{parisi2020quantum}. Interestingly, the MGP starts to underestimate $\rho^{(1)}(x)$ for larger values of $\delta g/g$ as the particle number is decreased. Herein, this critical value is $\delta g /g \approx 0.4$ for $N=40$ and $\delta g/g\approx0.5$ for $N=10$. Moreover, independently of the value of $\delta g/g$ an increasing atom number leads to a transition from a Gaussian-like to a FT profile and then to an almost homogeneous density distribution within the MGP method. 
The homogeneous density peak is determined in the box potential by $\max [N/L, n_{0}]$, see also Eq.~(\ref{equil_den}). 
The relevant atom number to realize such  drastic deformations is relatively large and thus being out of reach of the MB approach. 
For instance, considering a box potential of length $L=800$, in the case of $\delta g/g=0.2$ we observe a FT for $N>80$ while a homogeneous profile is attained for $N>500$ [Fig.~\ref{fig:den_sym_varydg} (e)]. 
We note that for smaller interactions, e.g. $\delta g/g=0.02$ the corresponding FT (homogeneous) behavior is achieved for $N>2000$ ($N>50000$), see also~\footnote{The crossover behavior from a Gaussian-shaped to a FT and then to a homogeneous profile for a specific $\delta g /g$ and increasing atom number occurs 
independently of the finite size of the system, i.e. the length of the 
box potential. In particular the relevant particle number to realize this transition is smaller for decreasing $L$. 
For example, within MGP, if $L=100$ and $g=0.05$ then for $\delta g/g=0.02$ [$\delta g/g=0.2$] we achieve a FT when $N>2000$ [$N>100$] and a homogeneous distribution for $N>10^5$ [$N>5000$].}. 

\begin{figure*}[ht]
\includegraphics[width=0.9\textwidth]{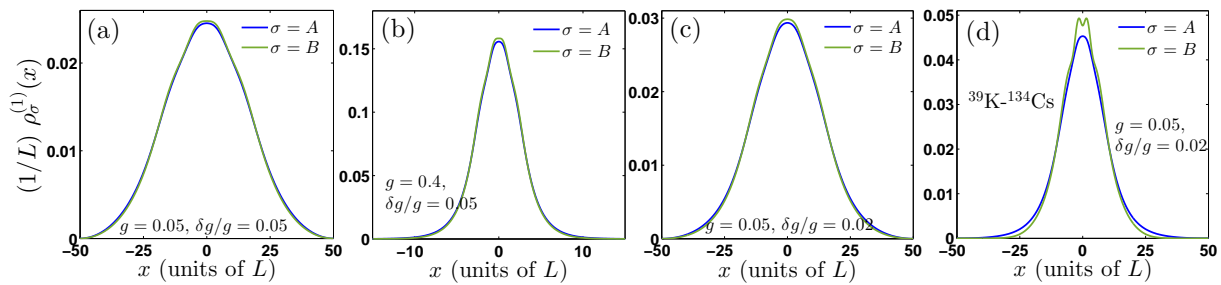}
\caption{Ground state densities $\rho^{(1)}_{\sigma}(x)$ of a mass-imbalanced mixture ($M_B \approx 2.12 M_A$) for (a), (b) distinct intracomponent coefficients $g$ and fixed $\delta g/ g=0.05$ (see legends) and (c) $\delta g/g=0.02$, $g=0.05$. 
The density profiles become narrower (flattened) for an increasing $g$ ($\delta g /g$).
(d) $\rho^{(1)}_{\sigma}(x)$ for a strongly mass-imbalanced setting with $M_B \approx 19 M_A$ while $\delta g/g=0.02$ and $g=0.05$. 
Excitation signatures are captured in the distribution of the heavy component being identified by the deformations of its respective density.  
Due to the mass-imbalance the components are discernible (mixed). 
The mixture has $N_A=N_B=20$ bosons residing in a box trap of length $L$. }
\label{fig:den_imbalance_varyg} 
\end{figure*}

\subsection{Effect of the harmonic trap}\label{trap}

Having exemplified the properties and interaction dependence of the droplet configurations emerging in symmetric mixtures which experience a box potential we next briefly study their formation in the presence of an external harmonic trap. 
In this case the MB Hamiltonian of the system is described by Eq.~(\ref{eq:MB_Hamilt})  with the addition of the external confinement $V(x)=(1/2)M \omega^2 x^2$. 
The latter naturally sets an additional length scale into the system, namely the harmonic oscillator length $a_{ho}=\sqrt{\hbar/M\omega} \approx 3.16$ with $\omega=0.1$ corresponding
to a trap of strength $2 \pi \times 20~Hz$ being an experimentally customary~\cite{wenz2013few,katsimiga2020observation} trap frequency in the longitudinal direction. As a consequence, the corresponding density distributions shrink as compared to the case of a box potential~\cite{cui2021droplet,cheiney2018bright}. As such the underlying densities feature mainly a Gaussian-shaped configuration whose width becomes narrower for increasing (decreasing) $g$ ($\delta g/g$) and keeping $\delta g/g$ ($g$) constant, see Figs.~\ref{fig:trap_den_sym}(a), (b) when $N=40$. 
The deviations between the MGP and MB approach in this trapped case become substantial as shown exemplarily in Fig.~\ref{fig:trap_den_sym}(c). 
Indeed, the density as estimated through MGP is narrower compared to the MB case which also shows FT signatures. 
This difference is more pronounced for either increasing trapping frequency or interaction parameter $g$ where trap effects are 
more transparent.
This is an expected result since MGP is constructed within the local density approximation~\cite{petrov2016ultradilute}.

The effect of the trap frequency for a specific $g$ and $\delta g/g$ is depicted in Fig.~\ref{fig:trap_den_sym}(c). 
As anticipated, $\rho^{(1)}(x)$ is wider for a smaller $\omega$ since $a_{ho}$ becomes larger. 
Relying on the MGP approach and for large particle numbers e.g. $N\sim 10^3$ we are able to realize an arguably flattened configuration e.g. for $\delta g/g=0.05$, $g=0.5$ and $\omega=0.05$. 
A further increase of the atom number, e.g. $N\sim 10^4$, results in a gradual delocalization of the density being reminiscent of a FT profile (not shown). 
For smaller trap frequencies, i.e. $\omega<0.01$, where trap effects are minimized droplets with a FT shape can be formed.   
In this way, it would be intriguing in the future to carefully inspect the interplay of the interaction coefficients, the particle number and the external confinement strength as well as to describe the transition from solitonic configurations to droplets in the region $\delta g/g<0$ as was recently discussed in Ref.~\cite{cui2021droplet} in a quasi-1D geometry. 
A similar delocalization tendency occurs upon considering an increasing atom number $N$ whilst all other system parameters are kept fixed, see Fig.~\ref{fig:trap_den_sym} (d).

\section{Droplets in mass-imbalanced mixtures}\label{assymetric_drops} 

Another far less investigated situation is the formation
of droplets in mass-imbalanced mixtures. 
This is a characteristic case example where the participating components are not identical and thus complicated mixed phases characterized by different degrees of miscibility may emerge~\cite{naidon2021mixed}. 
Recall that the reduction of the full two-component system to an effectively single-component one at the MGP level is no longer valid for mass-imbalanced setups. 
Relevant extensions of the MGP approach have been explored in three-dimensions~\cite{minardi2019effective,burchianti2020dual} containing a more complex LHY correction than the symmetric case.   

To elaborate on the impact of the mass ratio among the components we employ the experimentally realized~\cite{d2019observation,burchianti2020dual} heteronuclear mixture of $^{41}$K and $^{87}$Rb. 
Generally, the response of the mixture (at least on the single-particle level) to variations of the involved interaction strengths is qualitatively similar to the mass-balanced case, see Figs.~\ref{fig:den_imbalance_varyg} (a)-(c). 
However, the two components become discernible and for either increasing $g$ and keeping $\delta g/g$ fixed or vice versa their differentiation is less pronounced with the heavier component (immobile) exhibiting a narrower distribution than the lighter one~\cite{maity2020parametrically}. 
More concretely, for increasing $g$ while $\delta g/g$ is held constant the density distributions of both components become gradually narrower [Figs.~\ref{fig:den_imbalance_varyg}(a), (b)]. 
In the reverse case, where $\delta g/g$ is increased 
(while $g$ remains fixed), both $\rho_A^{(1)}(x)$ and $\rho_B^{(1)}(x)$ are slightly flattened, compare Figs.~\ref{fig:den_imbalance_varyg} (a), (c). 
The mixed character of the components can be readily enhanced by considering a larger mass ratio as depicted in Fig.~\ref{fig:den_imbalance_varyg}(d) regarding a mixture of $^{39}$K and $^{134}$Cs. 
Interestingly, $\rho_B^{(1)}(x)$ of the $^{134}$Cs component shows a two-hump structure signalling the excited nature of this subsystem and thus of the ensuing droplet configuration. 
This is a manifestation of the participation of higher-order correlations associated with the non-negligible population of higher-lying orbitals [Eq.~(\ref{eq:wfn_numb_states})] where such two-hump structures build upon. 
In this sense the excited nature of the system essentially refers to the admixing or otherwise fragmented character of the total MB state. 
Similar effects can also be observed for $^{174}$Yb (not shown). 
In that light, an interesting prospect for future studies is to carefully unveil possible excitation mechanisms of droplets arising in strongly mass-imbalanced mixtures and the further 
impact of such imbalance in direct numerical simulations.
\begin{figure*}[ht]
\includegraphics[width=0.9\textwidth]{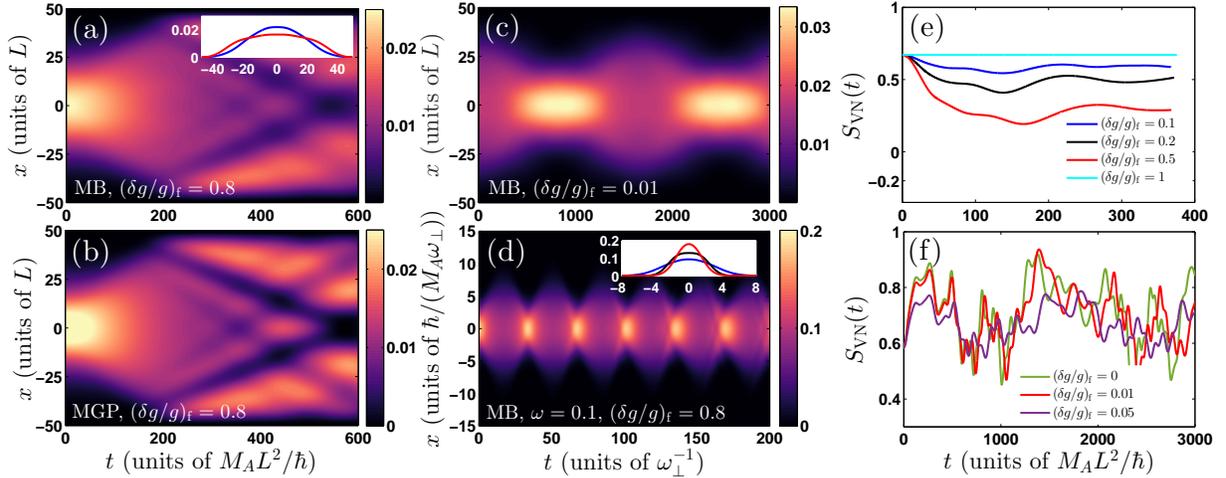}
\caption{Time-evolution of the single-particle density $\rho^{(1)}(x;t)$ of a box trapped symmetric bosonic mixture upon considering an interaction quench from $(\delta g/ g)_{\rm {in}}=0.02$ to $(\delta g/ g)_{\rm {f}}=0.8$ within (a) the MB and (b) the MGP framework. 
An expansion of the droplet, due to the reduction of the intercomponent attraction caused by the quench, towards the box edges occurs 
and is found to be less pronounced in the MB scenario. Inset in (a) presents density profiles at $t=50$ (blue line) and $t=150$ (red line) where FT signatures are evident. 
(c) The same as (a) but for a quench from $(\delta g/ g)_{\rm {in}}=0.2$ to $(\delta g/ g)_{\rm {f}}=0.01$. 
The mixture tends towards an LHY fluid due to the quench and it undergoes a breathing motion. 
The postquench intercomponent attraction is larger compared to its prequench value. 
(d) Same as (a) but for a harmonically trapped mixture. Breathing of the droplet is observed due to the existence of the external trap. 
The inset in (d) shows characteristic instantaneous density profiles depicting expansion $t=53$ (blue line), contraction $t=68$ (red line) and FT formation in between $t=60$ (black line). 
Dynamics of the von-Neumann entropy for different postquench interaction strengths (see legend) starting from (e) $(\delta g/ g)_{\rm {in}}=0.02$ and (f) $(\delta g/ g)_{\rm {in}}=0.2$. A dynamical reduction (enhancement) of the intercomponent entanglement is triggered for larger (smaller) postquench values of $\delta g/g$. 
The mixture consists of $N=40$ bosons of the same mass and intracomponent interactions $g_{AA}=g_{BB} \equiv g= 0.05$ for the box while $g_{AA}=g_{BB} \equiv g=0.1$ for the trap.}
\label{fig:dynamics_den_sym} 
\end{figure*}

\section{Interaction quench dynamics of droplets}\label{quenches}

After analyzing in detail the ground state of symmetric and asymmetric droplets, a natural next step is to investigate their dynamical behavior aiming in particular to elucidate their inherent build-up of correlations. 
To trigger the time-evolution of these structures we rely on an intercomponent interaction quench protocol associated with a sudden change of the $\delta g/g$ parameter while letting $g$ intact. 
Concretely, below, we consider quenches from small to larger values of $\delta g/g$ and vice versa for both homogeneously and harmonically trapped symmetric mixtures. 
Afterwards, we briefly comment on the dynamics of mass-imbalanced settings in a box potential.  
\begin{figure*}[ht]
\includegraphics[width=0.8\textwidth]{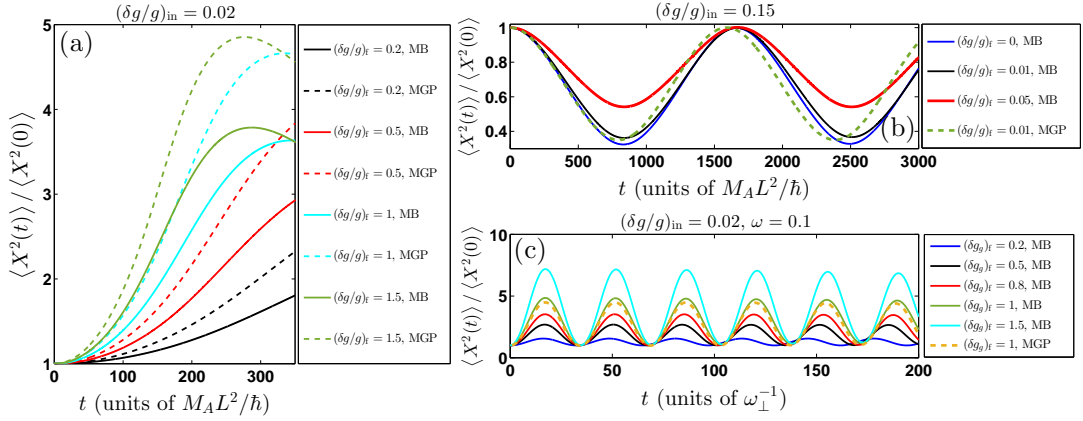}
\caption{Dynamics of the variance after an interaction quench of a symmetric bosonic mixture experiencing (a), (b) a box potential of length $L$ and (c) a harmonic trap of $\omega=0.1$ as obtained within the MB and the MGP approaches (see legends). 
The behavior of the variance quantifies (a) the expansion and (b), (c) the breathing dynamics of the mixture. Beyond LHY correlations lead to a pronounced reduction of the expansion rate (panel (a)) and a slight negative shift in the breathing frequency of the droplet (panels (c), (d)). 
The values of the pre- and postquench interactions $(\delta g/ g)_{\rm {in}}$ and $(\delta g/ g)_{\rm {f}}$ are shown in the legends. 
The bosonic mixture has $N=40$ atoms with equal mass and intracomponent interactions $g_{AA}=g_{BB} \equiv g =0.05$.}
\label{fig:variance_sym} 
\end{figure*}

\subsection{Box potential}\label{box_quench}

It can be easily deduced that upon quenching to a larger $(\delta g /g)_{\rm{f}}$, as compared to the initial one $(\delta g /g)_{\rm{in}}$, leads to relatively weaker postquench intercomponent attractions. 
Thus, a dynamical expansion of the bosonic cloud is anticipated. 
The resulting single-particle density of the mixture in the course of the evolution is provided in Fig.~\ref{fig:dynamics_den_sym} (a), (b) within the MB and the MGP approaches respectively for a quench characterized by $(\delta g /g)_{\rm{in}}=0.02$ and $(\delta g /g)_{\rm{f}}=0.8$. As it can be seen, the initial ($t=0$) Gaussian-shape profile of $\rho^{(1)}(x)$ due to $(\delta g /g)_{\rm{in}}=0.02$ undergoes an expansion towards the box edges. 
In the process ($t\sim 60$), it acquires a shape
reminiscent of a FT (see the inset of Fig.~\ref{fig:dynamics_den_sym} (a)) and upon hitting to the edges of the box potential ($t\sim 200$) it is reflected back to the center ($x=0$) while splitting into two parts. 
Namely, an inner portion and two outer humps close to the edges with the former traveling faster to $x=0$ and featuring later on several interference fringes after colliding with the latter humps. 
The above-described overall dynamical deformation of $\rho^{(1)}(x)$ takes place both in the MB and the MGP approaches but in the latter case the expansion strength and velocity of the bosonic cloud are substantially larger. 
As a result also the interference wave phenomenon appearing for long evolution times is more pronounced in the MGP dynamics.  

To estimate the deviation in the quench-induced expansion of the mixture between the MB and the MGP methods we resort to the position variance of the bosonic cloud in the course of the evolution given by
\begin{equation} 
\braket{X_{\sigma}^2(t)}=\braket{\Psi(t)|\hat{x}_{\sigma}^2|\Psi(t)}.\label{eq:variance} 
\end{equation}
Here, we have taken into account that $\braket{\Psi(t)|\hat{x}_{\sigma}|\Psi(t)}=0$. For symmetric mixtures it holds that $\braket{X_{A}^2(t)}=\braket{X_{B}^2(t)}\equiv \braket{X^2(t)}$. Monitoring $\braket{X^2(t)}/\braket{X^2(0)}$ for quenches from $(\delta g /g)_{\rm{in}}=0.02$ to various values of $(\delta g /g)_{\rm{f}}$ [Fig.~\ref{fig:variance_sym}(a)] we observe that it exhibits an increasing behavior quantifying expansion independently of $(\delta g /g)_{\rm{f}}$. 
Furthermore, the expansion amplitude and velocity ($\sim \delta \braket{X^2(t)}/\delta t$) become more enhanced for a larger $(\delta g /g)_{\rm{f}}$. 
Importantly, a comparison of $\braket{X^2(t)}/\braket{X^2(0)}$ for fixed $(\delta g /g)_{\rm{f}}$ among the different approaches reveals that both the strength and the velocity of the expansion are overestimated in the MGP dynamics. 
This is a signature of substantial beyond LHY correlations that build-up during the time-evolution. 
Since this observable is experimentally tractable via {\it in-situ} absorption imaging~\cite{fukuhara2013quantum,ronzheimer2013expansion}, it can be used as a probe for the presence of beyond LHY correlations during the expansion dynamics of droplet structures. 
\begin{figure*}[ht]
\includegraphics[width=0.9\textwidth]{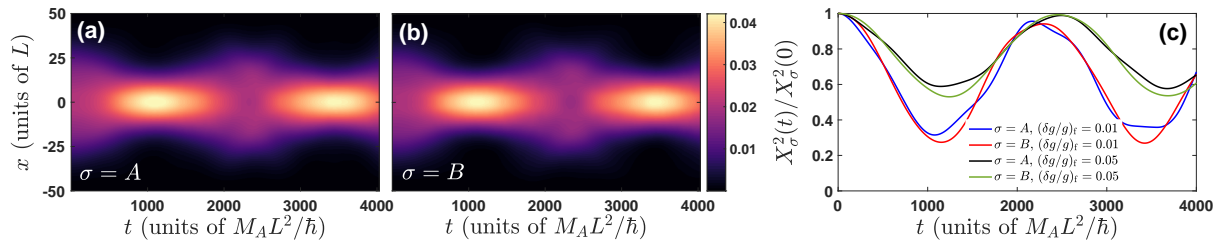}
\caption{Quenched $\sigma$-component density $\rho^{(1)}_{\sigma}(x)$ of a mass-imbalanced mixture after a change from $(\delta g/ g)_{\rm {in}}=0.15$ to $(\delta g/ g)_{\rm {f}}=0.01$ within the MB approach. 
(c) Time-evolution of the corresponding variances. 
The components perform a breathing dynamics showing a phase difference with the heavy component being delayed. 
The box trapped mass-imbalanced mixture has $N=40$ bosons with $M_B=2.1M_A$ and interactions $g_{AA}=g_{BB} \equiv g= 0.05$.}
\label{fig:dynamics_den_imbalance} 
\end{figure*} 

Additionally, a measure that has been so far elusive for droplets concerns their degree of entanglement (intercomponent correlations) which cannot be taken into account, at least properly, within the widely used MGP approach [Eq.~\ref{eq:MGP_asym}]. 
In order to quantify the amount of entanglement we herein rely on the von-Neumann entropy~\cite{horodecki2009quantum} which is defined as 
\begin{align}
S_{\rm{VN}}(t)=-\sum\limits_{k=1}^D \lambda_{k}(t)\ln[\lambda_k(t)].\label{eq:VN} 
\end{align} 
with $\lambda_k(t)$ being the Schmidt coefficients introduced in Eq.~(\ref{eq:wfn_total}). 
They are the eigenvalues of the $\sigma$-component reduced density matrix
$\rho_{\sigma}^{N_{\sigma}} (\vec{x}, \vec{x}';t)= \braket{\Psi(t)|\prod_{i=1}^{N_{\sigma}} \Psi_{\sigma}^{\dagger} (x_i) \prod_{i=1}^{N_{\sigma}} \Psi_{\sigma}(x_i')|\Psi(t)} $, where $\vec{x}=(x_1, \cdots,x_{N_{\sigma}})$. 
Evidently, if $S_{\rm{VN}}(t)\neq0$ implies that more than one  Schmidt coefficients contribute to the MB wave function of Eq.~(\ref{eq:wfn_total}) and hence the mixture is entangled. 
The time-evolution of $S_{\rm{VN}}(t)$ following a quench from $(\delta g /g)_{\rm{in}}=0.02$ to different but larger $(\delta g /g)_{\rm{f}}$ is showcased in Fig.~\ref{fig:dynamics_den_sym}(e). 
Notice that $S_{\rm{VN}}(0)$ is finite implying that the initial state (ground state droplet) is already entangled. Moreover, $S_{\rm{VN}}(t)$ reduces as time-evolves and in particular when the cloud expands and after the collision with the box edges it tends to saturate towards a $(\delta g /g)_{\rm{f}}$-dependent finite value. 
We remark that the reduction of $S_{\rm{VN}}(t)$ is caused by the mere fact that the postquench values refer to smaller $\abs{g_{AB}}$ couplings~\cite{morera2020quantum} while $S_{\rm{VN}}(t)$ is constant for $(\delta g /g)_{\rm{f}}=1$ since then $g_{AB}=0$. 

In the reverse quench scenario, where $(\delta g /g)_{\rm{f}}$ is smaller than $(\delta g /g)_{\rm{in}}$, the postquench $g_{AB}$ becomes more attractive and the gas approaches dynamically an LHY fluid. 
This protocol results in the dynamical contraction of $\rho^{(1)}(x;t)$ and ultimately triggers the breathing motion of the cloud~\cite{astrakharchik2018dynamics}, see for instance Fig.~\ref{fig:dynamics_den_sym}(c) corresponding to $(\delta g /g)_{\rm{f}}=0.01$. 
The emergent breathing dynamics can be readily captured by $\braket{X^2(t)}/\braket{X^2(0)}$ performing oscillations [Fig.~\ref{fig:variance_sym}(b)] and thus materializing the underlying expansion and contraction of the droplet. 
The oscillation amplitude is enhanced for $(\delta g /g)_{\rm{f}}\to 0$, while it is always slightly reduced in the MB case as compared to the MGP evolution. 
The deviation in the oscillation amplitude between the two approaches is typically of the order of $\sim 3.2\%$. 
Furthermore, the oscillation frequency of $\braket{X^2(t)}/\braket{X^2(0)}$ which refers to the breathing frequency of the system is smaller in the MB dynamics as compared to the MGP one. 
For instance $\omega_{br}\approx 0.0036$ in the MB case and $\omega_{br}\approx 0.004$ within the MGP framework, a result that holds irrespectively of $(\delta g /g)_{\rm{f}}$, see also Ref.~\footnote{The breathing frequency for fixed $\delta g/g$ and $g$ shows a weakly decreasing tendency for a smaller particle number in accordance to the observations made in Ref.~\cite{parisi2020quantum,astrakharchik2018dynamics}.}. 
Interestingly, a dephasing of $\braket{X^2(t)}/\braket{X^2(0)}$ occurs within the MB scenario being absent in the MGP approach, see also remark~\footnote{Notice that we do not observe any appreciable signatures of dephasing in the dynamics of $\braket{X^2(t)}$ within the MGP approach at least 
for total evolution times $T \leq20000$. This statement holds irrespectively of the postquench value e.g. $(\delta g/g)_{\rm f}=0,0.01,0.05$ and particle numbers $N=10,20,40,100$.}. This phenomenon constitutes another imprint of beyond LHY contributions in the course of the evolution. 
To further support this argument we provide $S_{\rm{VN}}(t)$ in Fig.~\ref{fig:dynamics_den_sym} (f). 
At the initial stages of the dynamics it features an overall increasing tendency associated with the build-up of intercomponent entanglement during contraction. 
Later on, it exhibits a multifrequency oscillatory behavior without any clear hierarchy in terms of $(\delta g/g)_{\rm f}$. 
\begin{figure*}[ht]
\includegraphics[width=0.8\textwidth]{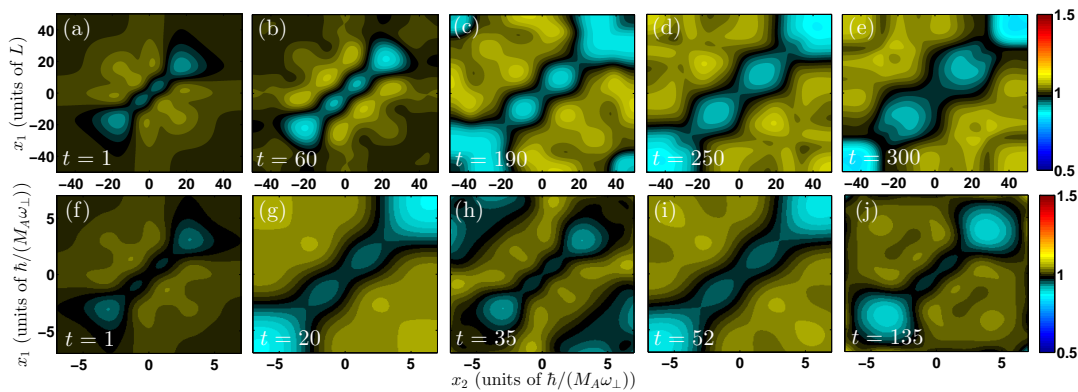}
\caption{Snapshots of the two-body correlation function following an interaction quench from $(\delta g/ g)_{\rm {in}}=0.02$ to $(\delta g/ g)_{\rm {f}}=0.8$ in (a)-(e) box potential of length $L$ and (f)-(j) harmonic trap with $\omega=0.1$. 
Anti-correlations develop at the same location of the droplet and a correlated behavior appears at distinct places. The symmetric mixture consists of $N=40$ mass-balanced bosons and intracomponent interactions (a)-(e) $g_{AA}=g_{BB}\equiv g=0.05$ whilst (f)-(j) $g_{AA}=g_{BB}\equiv g=0.1$.}
\label{fig:2bcoherence_sym} 
\end{figure*}

\subsection{Harmonic trap}\label{trap_quench}

In the presence of a harmonic trap the quench, independently of being performed to weaker or stronger $(\delta g /g)_{\rm{f}}$, leads inevitably to a collective breathing motion of the droplet. 
A characteristic case example of the ensuing density evolution is illustrated in Fig.~\ref{fig:dynamics_den_sym}(d) following a quench from $(\delta g /g)_{\rm{in}}= 0.02$ to $(\delta g /g)_{\rm{f}}= 0.8$. 
Interestingly, during the dynamics $\rho^{(1)}(x;t)$ shows signatures of a flattened distribution between full expansion and contraction while it exhibits a clear Gaussian shape being wider at the expansion than the contraction points, see also the inset of Fig.~\ref{fig:dynamics_den_sym} (d).   
As expected, this breathing motion is captured by the oscillatory motion of $\braket{X^2(t)}/\braket{X^2(0)}$ whose amplitude is larger for increasing postquench parameter. 
The MGP approach describes adequately both the density shape and the breathing frequency of the cloud, see e.g. Fig.~\ref{fig:variance_sym}(c). 
It predicts a slightly narrower $\rho^{(1)}(x;t)$ and larger breathing frequency. The latter being $\omega_{br} \approx 1.84 \omega$ in the MB case and $\omega_{br}\approx 1.85 \omega$ in the MGP scenario when $(\delta g /g)_{\rm{f}}= 0.8$. 
We note that $\omega_{br}$ appears to be almost insensitive to the value of $(\delta g /g)_{\rm{f}}$. 
For instance, regarding quenches to weaker interactions, e.g. from $(\delta g /g)_{\rm{in}}= 0.15$ to $(\delta g /g)_{\rm{f}}=0.01$, that we have found that it changes slightly having a value $\omega_{br} \approx 2 \omega$ [$\omega_{br}\approx 2.1 \omega$] in the MB [MGP] scenario.

\subsection{Mass-imbalanced mixture in the box}\label{mass_quench}

Next, we briefly analyze the quench dynamics of a mass-imbalanced mixture following an abrupt change to stronger intercomponent attractions such that we approach the MF balance point ($\delta g=0$) where an LHY fluid~\cite{jorgensen2018dilute,skov2021observation} forms, see Fig.~\ref{fig:dynamics_den_imbalance}. As explicated for the mass-balanced case [Fig.~\ref{fig:dynamics_den_sym}(c)] the quench causes  the two-component gas to perform a breathing motion. 
However, in this mass-imbalanced scenario due to the initial miscible
nature of the system [Fig.~\ref{fig:den_imbalance_varyg}(c)], the components oscillate with a phase difference between them which results into their slighly distinct breathing frequency. 
The latter is naturally smaller for the heavier component while the interspecies miscibility is retained throughout the evolution, see $\rho^{(1)}_A(x;t)$ and $\rho^{(1)}_B(x;t)$ in Fig.~\ref{fig:dynamics_den_imbalance}(a), (b). 
This observation is corroborated also by inspecting $\braket{X_{\sigma}^2(t)}/\braket{X_{\sigma}^2(0)}$ [Fig.~\ref{fig:dynamics_den_imbalance}(c)] whose oscillatory behavior reflects the expansion and contraction of the clouds. Distortions that are present in the oscillation of $\braket{X_{\sigma}^2(t)}/\braket{X_{\sigma}^2(0)}$ stem from the mixedness of the subsystems and their intercomponent interaction.
As for mass-balanced mixtures the oscillation amplitude of $\braket{X_{\sigma}^2(t)}/\braket{X_{\sigma}^2(0)}$ is larger when $(\delta g /g)_{\rm{f}}\to 0$. 
It is also worth mentioning that applying quenches to weaker $g_{AB}$ the clouds experience an expansion trend with a velocity smaller for the heavy component (not shown).

\subsection{Correlation patterns of droplets}\label{correl_dyn_drop}

As argued above, an important ingredient of the droplet formation is the presence of correlations which are expected to be especially enhanced during the nonequilibrium dynamics of such structures. 
Focusing on intercomponent entanglement we have indicated that besides being finite already in the ground state it is enhanced (reduced) after quenching to a weaker (stronger) $(\delta g /g)_{\rm{f}}$. 
Next, we aim at unveiling the emergent intracomponent dynamical correlation patterns of droplets. 
For this reason, we monitor the normalized spatially resolved two-body intracomponent correlation function~\cite{mistakidis2018correlation}
\begin{equation}
G^{(2)}_{\sigma\sigma}(x_1,x_2;t)=\frac{\rho_{\sigma\sigma}^{(2)}(x_1,x_2;t)}{\rho_\sigma^{(1)}(x_1;t)\rho_{\sigma}^{(1)}(x_2;t)}. \label{eq:two_body_coherence}
\end{equation} 
Importantly, it holds that if $G^{(2)}_{\sigma\sigma}(x_1,x_2;t)>1$ [$G^{(2)}_{\sigma\sigma}(x_1,x_2;t)<1$] the MB state is termed correlated [anti-correlated], whilst for $G^{(2)}_{\sigma\sigma}(x_1,x_2;t)=1$ it is fully second order coherent~\cite{mistakidis2018correlation}. 
This observable can be experimentally attained via {\it in-situ} density-density fluctuation 
measurements~\cite{nguyen2019parametric}. 
Subsequently we shall focus on the dynamics of symmetric mixtures and thus it holds that $G^{(2)}_{AA}(x_1,x_2;t)=G^{(2)}_{BB}(x_1,x_2;t)\equiv G^{(2)}(x_1,x_2;t)$. 
Also, we analyze $G^{(2)}(x_1,x_2;t)$ utilizing a quench from $(\delta g /g)_{\rm{in}}=0.02$ towards $(\delta g /g)_{\rm{f}}=0.8$ in a box potential [Figs.~\ref{fig:2bcoherence_sym}(a)-(e)] and in the presence of a trap [Fig.~\ref{fig:2bcoherence_sym}(f)-(j)]. 

Referring to a box potential and for this quench scenario the bosonic cloud experiences an overall expansion, see Fig.~\ref{fig:dynamics_den_sym}(a). 
The corresponding $G^{(2)}(x_1,x_2;t)$ is showcased in Figs.~\ref{fig:2bcoherence_sym}(a)-(e) at different time-instants of the evolution. 
As can be seen, two bosons of the same component feature an anti-correlated behavior at the same position (see the diagonal of $G^{(2)}(x_1,x_2=x_1)<1$) while they exhibit a correlation tendency when being far apart ($G^{(2)}(x_1,x_2\neq x_1)>1$). 
This pattern is maintained as time evolves and it is particularly enhanced as the cloud expands further. 

Turning to the dynamics in the presence of a harmonic trap, we observe the build-up of similar to the above-described correlation patterns, see Figs.~\ref{fig:2bcoherence_sym}(f)-(j). 
Recall that in this case the bosons perform a collective breathing motion [Fig.~\ref{fig:dynamics_den_sym}(d)] and characteristic snapshots of $G^{(2)}(x_1,x_2;t)$ are provided as examples herein during expansion [Figs.~\ref{fig:2bcoherence_sym}(g), (i)] and contraction [Figs.~\ref{fig:2bcoherence_sym} (h), (j)]. 
It can be readily deduced that two-body correlations exist for bosons that lie apart ($G^{(2)}(x_1,x_2\neq x_1;t)>1$) and anti-correlations persist in the same location ($G^{(2)}(x_1,x_2= x_1;t)<1$). 
Interestingly, two bosons show an anti-correlated tendency in the vicinity of the spatial region of their contraction, see e.g. $G^{(2)}(-3<x_1<3,-3<x_2<3;t)\approx 0$ in Figs.~\ref{fig:2bcoherence_sym} (h), (j).

\section{Summary and Outlook}\label{sec:conclusions}

We have investigated the impact of correlations on the ground state formation and quantum dynamics of droplet configurations arising in two-component one-dimensional particle-balanced bosonic settings experiencing either a box potential or a harmonic trap (for both mass-balanced and mass-imbalanced settings). 
The considered mixtures are characterized by the same intracomponent repulsion and an intercomponent attraction ensuring the generation of droplets. 
A particular focus is placed on the emergent droplet correlation patterns, the crossover from few- to many-atoms and the effect of the intra- and intercomponent interactions.  We thus explore
the most-general two-component setting even for the symmetric cases where according to the MGP framework the mixture reduces to a single component setup. 
The impact of the particle number in LHY fluids is also addressed. 
Importantly, the quench dynamics of droplets within a nonperturbative approach is explicated and deviations from the MGP predictions are exposed. Indeed, for the low number of atoms where the modified Gross-Pitaevskii and the MB approaches were compared, nontrivial differences in features such as the 
width and amplitude of the formed droplets were observed.

For the ground state of droplets we initially investigate the impact of an increasing intracomponent repulsion while adjusting the intercomponent attraction such that a suitable ratio remains constant. 
A transition from a spatially delocalized to a highly localized Gaussian density distribution takes place for stronger attractions irrespectively of the atom number. Signatures of a flattened density maximum are identified for intermediate interactions. 
Moreover, a larger particle number for fixed interactions results in the deformation of the droplet shape from a Gaussian-like towards a wider FT density profile. 
In sharp contrast, for suppressed MF interactions LHY fluids are realized experiencing a gradually shrinked configuration as the atom number increases. 
Simultaneously, on the two-body level a correlation hole appears for few-atom drops being suppressed for larger atom numbers giving its place to a tendency for long-range two-body correlations. 
Furthermore, by considering a fixed intracomponent repulsion and a decreasing intercomponent attraction a deformation from a Gaussian-shape to a pronounced FT distribution occurs. Remarkably, droplets with a FT exhibit a correlation hole which vanishes for localized ones. 

In the presence of a harmonic trap a crossover behavior from a delocalized to a strongly localized droplet shape occurs for either increasing intercomponent attraction or larger intracomponent repulsions. 
A tighter trap results into the suppression of any FT signature. 
In all cases, careful comparisons between the MB and other MF-type approaches operating at different correlation orders are performed. 
This way, we expose deviations in the droplet profile ultimately caused by the presence of intercomponent entanglement. 
These differences are maximized for stronger attractions among the species and also in the presence of a harmonic trap where the MGP is less adequate due to the local density approximation. 

Regarding weakly mass-imbalanced mixtures we illustrate
that a similar to the above-described structural deformation takes place upon varying the individual interaction parameters. 
Here, the components are mixed with respect to each other with the heavier one having a narrower distribution. 
Strikingly, an adequately heavy component signals the formation of a
multi-humped droplet structure. 

Next, we study the nonequilibrium dynamics of droplets upon applying intercomponent interaction quenches to either weaker or stronger attractions. 
In the former case the droplet expands reshaping from a Gaussian-like to a configuration that bears FT characteristics, whilst for quenches towards the MF balance point it performs a breathing motion. Naturally, an accompanying build-up (reduction) of the intercomponent entanglement occurs for quenches to larger (smaller) attractions. 
In both cases evident deviations of the MB from the MGP dynamics are present manifesting in a reduced expansion velocity of the droplet within the MB approach and also a smaller breathing frequency as compared to the MGP prediction. 
In the presence of a harmonic trap, the quench leads to a collective breathing dynamics whose frequency is overestimated in the MGP case. 
Monitoring the development of two-body correlations during the droplet evolution we exemplify that irrespectively of the quench two bosons feature an anti-correlation at the same position while being correlated when residing far apart. 

There are several intriguing research directions that can be pursued in future endeavors further extending the present findings. 
Indeed, a detailed investigation of the different collective excitations of droplets emerging in mass- as well as interaction-imbalanced mixtures and the comparison with the MGP predictions is of direct interest. 
Moreover, the study of correlated pattern formation and associated defect scaling e.g. in the context of the Kibble-Zurek mechanism when crossing the threshold of MF repulsion and attraction utilizing a time-dependent protocol would be intriguing. 
Certainly, the engineering of the magnetic properties of droplet structures constitutes another exciting perspective. 
Moreover, we have limited our considerations herein
to one-dimensional settings and the most prototypical of
coherent structures within this system. However, much of the
recent work has focused on two- and three-dimensions
\cite{PhysRevLett.123.133901,PhysRevLett.122.193902,kartashov2018three} and excited states, including droplet clusters, vortices
etc. It will be particularly relevant to extend the MB
considerations herein to such higher-dimensional 
settings and some of the associated states.

\section*{Acknowledgements} 
S. I. M.  gratefully acknowledges financial support in the framework of the Lenz-Ising Award of the University of Hamburg. 
H.R.S. and S. I. M.   acknowledge support from the NSF through a grant for ITAMP at Harvard University.
P.G.K.  is supported by the US National Science Foundation under Grants No. PHY-2110030.

\bibliography{droplets}{}

\begin{thebibliography}{78}%
\makeatletter
\providecommand \@ifxundefined [1]{%
 \@ifx{#1\undefined}
}%
\providecommand \@ifnum [1]{%
 \ifnum #1\expandafter \@firstoftwo
 \else \expandafter \@secondoftwo
 \fi
}%
\providecommand \@ifx [1]{%
 \ifx #1\expandafter \@firstoftwo
 \else \expandafter \@secondoftwo
 \fi
}%
\providecommand \natexlab [1]{#1}%
\providecommand \enquote  [1]{``#1''}%
\providecommand \bibnamefont  [1]{#1}%
\providecommand \bibfnamefont [1]{#1}%
\providecommand \citenamefont [1]{#1}%
\providecommand \href@noop [0]{\@secondoftwo}%
\providecommand \href [0]{\begingroup \@sanitize@url \@href}%
\providecommand \@href[1]{\@@startlink{#1}\@@href}%
\providecommand \@@href[1]{\endgroup#1\@@endlink}%
\providecommand \@sanitize@url [0]{\catcode `\\12\catcode `\$12\catcode
  `\&12\catcode `\#12\catcode `\^12\catcode `\_12\catcode `\%12\relax}%
\providecommand \@@startlink[1]{}%
\providecommand \@@endlink[0]{}%
\providecommand \url  [0]{\begingroup\@sanitize@url \@url }%
\providecommand \@url [1]{\endgroup\@href {#1}{\urlprefix }}%
\providecommand \urlprefix  [0]{URL }%
\providecommand \Eprint [0]{\href }%
\providecommand \doibase [0]{http://dx.doi.org/}%
\providecommand \selectlanguage [0]{\@gobble}%
\providecommand \bibinfo  [0]{\@secondoftwo}%
\providecommand \bibfield  [0]{\@secondoftwo}%
\providecommand \translation [1]{[#1]}%
\providecommand \BibitemOpen [0]{}%
\providecommand \bibitemStop [0]{}%
\providecommand \bibitemNoStop [0]{.\EOS\space}%
\providecommand \EOS [0]{\spacefactor3000\relax}%
\providecommand \BibitemShut  [1]{\csname bibitem#1\endcsname}%
\let\auto@bib@innerbib\@empty
\bibitem [{\citenamefont {Papp}\ \emph {et~al.}(2008)\citenamefont {Papp},
  \citenamefont {Pino}, \citenamefont {Wild}, \citenamefont {Ronen},
  \citenamefont {Wieman}, \citenamefont {Jin},\ and\ \citenamefont
  {Cornell}}]{papp2008bragg}%
  \BibitemOpen
  \bibfield  {author} {\bibinfo {author} {\bibfnamefont {S.~B.}\ \bibnamefont
  {Papp}}, \bibinfo {author} {\bibfnamefont {J.~M.}\ \bibnamefont {Pino}},
  \bibinfo {author} {\bibfnamefont {R.~J.}\ \bibnamefont {Wild}}, \bibinfo
  {author} {\bibfnamefont {S.}~\bibnamefont {Ronen}}, \bibinfo {author}
  {\bibfnamefont {C.~E.}\ \bibnamefont {Wieman}}, \bibinfo {author}
  {\bibfnamefont {D.~S.}\ \bibnamefont {Jin}}, \ and\ \bibinfo {author}
  {\bibfnamefont {E.~A.}\ \bibnamefont {Cornell}},\ }\href@noop {} {\bibfield
  {journal} {\bibinfo  {journal} {Phys. Rev. Lett.}\ }\textbf {\bibinfo
  {volume} {101}},\ \bibinfo {pages} {135301} (\bibinfo {year}
  {2008})}\BibitemShut {NoStop}%
\bibitem [{\citenamefont {Navon}\ \emph {et~al.}(2010)\citenamefont {Navon},
  \citenamefont {Nascimbene}, \citenamefont {Chevy},\ and\ \citenamefont
  {Salomon}}]{navon2010equation}%
  \BibitemOpen
  \bibfield  {author} {\bibinfo {author} {\bibfnamefont {N.}~\bibnamefont
  {Navon}}, \bibinfo {author} {\bibfnamefont {S.}~\bibnamefont {Nascimbene}},
  \bibinfo {author} {\bibfnamefont {F.}~\bibnamefont {Chevy}}, \ and\ \bibinfo
  {author} {\bibfnamefont {C.}~\bibnamefont {Salomon}},\ }\href@noop {}
  {\bibfield  {journal} {\bibinfo  {journal} {Science}\ }\textbf {\bibinfo
  {volume} {328}},\ \bibinfo {pages} {729} (\bibinfo {year}
  {2010})}\BibitemShut {NoStop}%
\bibitem [{\citenamefont {Navon}\ \emph {et~al.}(2011)\citenamefont {Navon},
  \citenamefont {Piatecki}, \citenamefont {G{\"u}nter}, \citenamefont {Rem},
  \citenamefont {Nguyen}, \citenamefont {Chevy}, \citenamefont {Krauth},\ and\
  \citenamefont {Salomon}}]{navon2011dynamics}%
  \BibitemOpen
  \bibfield  {author} {\bibinfo {author} {\bibfnamefont {N.}~\bibnamefont
  {Navon}}, \bibinfo {author} {\bibfnamefont {S.}~\bibnamefont {Piatecki}},
  \bibinfo {author} {\bibfnamefont {K.}~\bibnamefont {G{\"u}nter}}, \bibinfo
  {author} {\bibfnamefont {B.}~\bibnamefont {Rem}}, \bibinfo {author}
  {\bibfnamefont {T.~C.}\ \bibnamefont {Nguyen}}, \bibinfo {author}
  {\bibfnamefont {F.}~\bibnamefont {Chevy}}, \bibinfo {author} {\bibfnamefont
  {W.}~\bibnamefont {Krauth}}, \ and\ \bibinfo {author} {\bibfnamefont
  {C.}~\bibnamefont {Salomon}},\ }\href@noop {} {\bibfield  {journal} {\bibinfo
   {journal} {Phys. Rev. Lett.}\ }\textbf {\bibinfo {volume} {107}},\ \bibinfo
  {pages} {135301} (\bibinfo {year} {2011})}\BibitemShut {NoStop}%
\bibitem [{\citenamefont {Shin}\ \emph {et~al.}(2008)\citenamefont {Shin},
  \citenamefont {Schirotzek}, \citenamefont {Schunck},\ and\ \citenamefont
  {Ketterle}}]{shin2008realization}%
  \BibitemOpen
  \bibfield  {author} {\bibinfo {author} {\bibfnamefont {Y.-i.}\ \bibnamefont
  {Shin}}, \bibinfo {author} {\bibfnamefont {A.}~\bibnamefont {Schirotzek}},
  \bibinfo {author} {\bibfnamefont {C.~H.}\ \bibnamefont {Schunck}}, \ and\
  \bibinfo {author} {\bibfnamefont {W.}~\bibnamefont {Ketterle}},\ }\href@noop
  {} {\bibfield  {journal} {\bibinfo  {journal} {Phys. Rev. Lett.}\ }\textbf
  {\bibinfo {volume} {101}},\ \bibinfo {pages} {070404} (\bibinfo {year}
  {2008})}\BibitemShut {NoStop}%
\bibitem [{\citenamefont {Petrov}(2015)}]{petrov2015quantum}%
  \BibitemOpen
  \bibfield  {author} {\bibinfo {author} {\bibfnamefont {D.~S.}\ \bibnamefont
  {Petrov}},\ }\href {\doibase 10.1103/PhysRevLett.115.155302} {\bibfield
  {journal} {\bibinfo  {journal} {Phys. Rev. Lett.}\ }\textbf {\bibinfo
  {volume} {115}},\ \bibinfo {pages} {155302} (\bibinfo {year}
  {2015})}\BibitemShut {NoStop}%
\bibitem [{\citenamefont {Petrov}\ and\ \citenamefont
  {Astrakharchik}(2016)}]{petrov2016ultradilute}%
  \BibitemOpen
  \bibfield  {author} {\bibinfo {author} {\bibfnamefont {D.~S.}\ \bibnamefont
  {Petrov}}\ and\ \bibinfo {author} {\bibfnamefont {G.~E.}\ \bibnamefont
  {Astrakharchik}},\ }\href@noop {} {\bibfield  {journal} {\bibinfo  {journal}
  {Phys. Rev. Lett.}\ }\textbf {\bibinfo {volume} {117}},\ \bibinfo {pages}
  {100401} (\bibinfo {year} {2016})}\BibitemShut {NoStop}%
\bibitem [{\citenamefont {Cabrera}\ \emph {et~al.}(2018)\citenamefont
  {Cabrera}, \citenamefont {Tanzi}, \citenamefont {Sanz}, \citenamefont
  {Naylor}, \citenamefont {Thomas}, \citenamefont {Cheiney},\ and\
  \citenamefont {Tarruell}}]{cabrera2018quantum}%
  \BibitemOpen
  \bibfield  {author} {\bibinfo {author} {\bibfnamefont {C.~R.}\ \bibnamefont
  {Cabrera}}, \bibinfo {author} {\bibfnamefont {L.}~\bibnamefont {Tanzi}},
  \bibinfo {author} {\bibfnamefont {J.}~\bibnamefont {Sanz}}, \bibinfo {author}
  {\bibfnamefont {B.}~\bibnamefont {Naylor}}, \bibinfo {author} {\bibfnamefont
  {P.}~\bibnamefont {Thomas}}, \bibinfo {author} {\bibfnamefont
  {P.}~\bibnamefont {Cheiney}}, \ and\ \bibinfo {author} {\bibfnamefont
  {L.}~\bibnamefont {Tarruell}},\ }\href@noop {} {\bibfield  {journal}
  {\bibinfo  {journal} {Science}\ }\textbf {\bibinfo {volume} {359}},\ \bibinfo
  {pages} {301} (\bibinfo {year} {2018})}\BibitemShut {NoStop}%
\bibitem [{\citenamefont {Cheiney}\ \emph {et~al.}(2018)\citenamefont
  {Cheiney}, \citenamefont {Cabrera}, \citenamefont {Sanz}, \citenamefont
  {Naylor}, \citenamefont {Tanzi},\ and\ \citenamefont
  {Tarruell}}]{cheiney2018bright}%
  \BibitemOpen
  \bibfield  {author} {\bibinfo {author} {\bibfnamefont {P.}~\bibnamefont
  {Cheiney}}, \bibinfo {author} {\bibfnamefont {C.~R.}\ \bibnamefont
  {Cabrera}}, \bibinfo {author} {\bibfnamefont {J.}~\bibnamefont {Sanz}},
  \bibinfo {author} {\bibfnamefont {B.}~\bibnamefont {Naylor}}, \bibinfo
  {author} {\bibfnamefont {L.}~\bibnamefont {Tanzi}}, \ and\ \bibinfo {author}
  {\bibfnamefont {L.}~\bibnamefont {Tarruell}},\ }\href@noop {} {\bibfield
  {journal} {\bibinfo  {journal} {Phys. Rev. Lett.}\ }\textbf {\bibinfo
  {volume} {120}},\ \bibinfo {pages} {135301} (\bibinfo {year}
  {2018})}\BibitemShut {NoStop}%
\bibitem [{\citenamefont {Ferioli}\ \emph {et~al.}(2019)\citenamefont
  {Ferioli}, \citenamefont {Semeghini}, \citenamefont {Masi}, \citenamefont
  {Giusti}, \citenamefont {Modugno}, \citenamefont {Inguscio}, \citenamefont
  {Gallem{\'\i}}, \citenamefont {Recati},\ and\ \citenamefont
  {Fattori}}]{ferioli2019collisions}%
  \BibitemOpen
  \bibfield  {author} {\bibinfo {author} {\bibfnamefont {G.}~\bibnamefont
  {Ferioli}}, \bibinfo {author} {\bibfnamefont {G.}~\bibnamefont {Semeghini}},
  \bibinfo {author} {\bibfnamefont {L.}~\bibnamefont {Masi}}, \bibinfo {author}
  {\bibfnamefont {G.}~\bibnamefont {Giusti}}, \bibinfo {author} {\bibfnamefont
  {G.}~\bibnamefont {Modugno}}, \bibinfo {author} {\bibfnamefont
  {M.}~\bibnamefont {Inguscio}}, \bibinfo {author} {\bibfnamefont
  {A.}~\bibnamefont {Gallem{\'\i}}}, \bibinfo {author} {\bibfnamefont
  {A.}~\bibnamefont {Recati}}, \ and\ \bibinfo {author} {\bibfnamefont
  {M.}~\bibnamefont {Fattori}},\ }\href@noop {} {\bibfield  {journal} {\bibinfo
   {journal} {Phys. Rev. Lett.}\ }\textbf {\bibinfo {volume} {122}},\ \bibinfo
  {pages} {090401} (\bibinfo {year} {2019})}\BibitemShut {NoStop}%
\bibitem [{\citenamefont {D'Errico}\ \emph {et~al.}(2019)\citenamefont
  {D'Errico}, \citenamefont {Burchianti}, \citenamefont {Prevedelli},
  \citenamefont {Salasnich}, \citenamefont {Ancilotto}, \citenamefont
  {Modugno}, \citenamefont {Minardi},\ and\ \citenamefont
  {Fort}}]{d2019observation}%
  \BibitemOpen
  \bibfield  {author} {\bibinfo {author} {\bibfnamefont {C.}~\bibnamefont
  {D'Errico}}, \bibinfo {author} {\bibfnamefont {A.}~\bibnamefont
  {Burchianti}}, \bibinfo {author} {\bibfnamefont {M.}~\bibnamefont
  {Prevedelli}}, \bibinfo {author} {\bibfnamefont {L.}~\bibnamefont
  {Salasnich}}, \bibinfo {author} {\bibfnamefont {F.}~\bibnamefont
  {Ancilotto}}, \bibinfo {author} {\bibfnamefont {M.}~\bibnamefont {Modugno}},
  \bibinfo {author} {\bibfnamefont {F.}~\bibnamefont {Minardi}}, \ and\
  \bibinfo {author} {\bibfnamefont {C.}~\bibnamefont {Fort}},\ }\href@noop {}
  {\bibfield  {journal} {\bibinfo  {journal} {Phys. Rev. Research}\ }\textbf
  {\bibinfo {volume} {1}},\ \bibinfo {pages} {033155} (\bibinfo {year}
  {2019})}\BibitemShut {NoStop}%
\bibitem [{\citenamefont {Burchianti}\ \emph {et~al.}(2020)\citenamefont
  {Burchianti}, \citenamefont {D’Errico}, \citenamefont {Prevedelli},
  \citenamefont {Salasnich}, \citenamefont {Ancilotto}, \citenamefont
  {Modugno}, \citenamefont {Minardi},\ and\ \citenamefont
  {Fort}}]{burchianti2020dual}%
  \BibitemOpen
  \bibfield  {author} {\bibinfo {author} {\bibfnamefont {A.}~\bibnamefont
  {Burchianti}}, \bibinfo {author} {\bibfnamefont {C.}~\bibnamefont
  {D’Errico}}, \bibinfo {author} {\bibfnamefont {M.}~\bibnamefont
  {Prevedelli}}, \bibinfo {author} {\bibfnamefont {L.}~\bibnamefont
  {Salasnich}}, \bibinfo {author} {\bibfnamefont {F.}~\bibnamefont
  {Ancilotto}}, \bibinfo {author} {\bibfnamefont {M.}~\bibnamefont {Modugno}},
  \bibinfo {author} {\bibfnamefont {F.}~\bibnamefont {Minardi}}, \ and\
  \bibinfo {author} {\bibfnamefont {C.}~\bibnamefont {Fort}},\ }\href@noop {}
  {\bibfield  {journal} {\bibinfo  {journal} {Condensed Matter}\ }\textbf
  {\bibinfo {volume} {5}},\ \bibinfo {pages} {21} (\bibinfo {year}
  {2020})}\BibitemShut {NoStop}%
\bibitem [{\citenamefont {Ferrier-Barbut}\ \emph {et~al.}(2016)\citenamefont
  {Ferrier-Barbut}, \citenamefont {Kadau}, \citenamefont {Schmitt},
  \citenamefont {Wenzel},\ and\ \citenamefont {Pfau}}]{ferrier2016observation}%
  \BibitemOpen
  \bibfield  {author} {\bibinfo {author} {\bibfnamefont {I.}~\bibnamefont
  {Ferrier-Barbut}}, \bibinfo {author} {\bibfnamefont {H.}~\bibnamefont
  {Kadau}}, \bibinfo {author} {\bibfnamefont {M.}~\bibnamefont {Schmitt}},
  \bibinfo {author} {\bibfnamefont {M.}~\bibnamefont {Wenzel}}, \ and\ \bibinfo
  {author} {\bibfnamefont {T.}~\bibnamefont {Pfau}},\ }\href@noop {} {\bibfield
   {journal} {\bibinfo  {journal} {Phys. Rev. Lett.}\ }\textbf {\bibinfo
  {volume} {116}},\ \bibinfo {pages} {215301} (\bibinfo {year}
  {2016})}\BibitemShut {NoStop}%
\bibitem [{\citenamefont {Chomaz}\ \emph {et~al.}(2016)\citenamefont {Chomaz},
  \citenamefont {Baier}, \citenamefont {Petter}, \citenamefont {Mark},
  \citenamefont {W{\"a}chtler}, \citenamefont {Santos},\ and\ \citenamefont
  {Ferlaino}}]{chomaz2016quantum}%
  \BibitemOpen
  \bibfield  {author} {\bibinfo {author} {\bibfnamefont {L.}~\bibnamefont
  {Chomaz}}, \bibinfo {author} {\bibfnamefont {S.}~\bibnamefont {Baier}},
  \bibinfo {author} {\bibfnamefont {D.}~\bibnamefont {Petter}}, \bibinfo
  {author} {\bibfnamefont {M.}~\bibnamefont {Mark}}, \bibinfo {author}
  {\bibfnamefont {F.}~\bibnamefont {W{\"a}chtler}}, \bibinfo {author}
  {\bibfnamefont {L.}~\bibnamefont {Santos}}, \ and\ \bibinfo {author}
  {\bibfnamefont {F.}~\bibnamefont {Ferlaino}},\ }\href@noop {} {\bibfield
  {journal} {\bibinfo  {journal} {Phys. Rev. X}\ }\textbf {\bibinfo {volume}
  {6}},\ \bibinfo {pages} {041039} (\bibinfo {year} {2016})}\BibitemShut
  {NoStop}%
\bibitem [{\citenamefont {Huang}\ and\ \citenamefont
  {Yang}(1957)}]{huang1957quantum}%
  \BibitemOpen
  \bibfield  {author} {\bibinfo {author} {\bibfnamefont {K.}~\bibnamefont
  {Huang}}\ and\ \bibinfo {author} {\bibfnamefont {C.~N.}\ \bibnamefont
  {Yang}},\ }\href@noop {} {\bibfield  {journal} {\bibinfo  {journal} {Phys.
  Rev.}\ }\textbf {\bibinfo {volume} {105}},\ \bibinfo {pages} {767} (\bibinfo
  {year} {1957})}\BibitemShut {NoStop}%
\bibitem [{\citenamefont {Luo}\ \emph {et~al.}(2021)\citenamefont {Luo},
  \citenamefont {Pang}, \citenamefont {Liu}, \citenamefont {Li},\ and\
  \citenamefont {Malomed}}]{luo2021new}%
  \BibitemOpen
  \bibfield  {author} {\bibinfo {author} {\bibfnamefont {Z.-H.}\ \bibnamefont
  {Luo}}, \bibinfo {author} {\bibfnamefont {W.}~\bibnamefont {Pang}}, \bibinfo
  {author} {\bibfnamefont {B.}~\bibnamefont {Liu}}, \bibinfo {author}
  {\bibfnamefont {Y.-Y.}\ \bibnamefont {Li}}, \ and\ \bibinfo {author}
  {\bibfnamefont {B.~A.}\ \bibnamefont {Malomed}},\ }\href@noop {} {\bibfield
  {journal} {\bibinfo  {journal} {Frontiers of Physics}\ }\textbf {\bibinfo
  {volume} {16}},\ \bibinfo {pages} {1} (\bibinfo {year} {2021})}\BibitemShut
  {NoStop}%
\bibitem [{\citenamefont {Kartashov}\ \emph {et~al.}(2018)\citenamefont
  {Kartashov}, \citenamefont {Malomed}, \citenamefont {Tarruell},\ and\
  \citenamefont {Torner}}]{kartashov2018three}%
  \BibitemOpen
  \bibfield  {author} {\bibinfo {author} {\bibfnamefont {Y.~V.}\ \bibnamefont
  {Kartashov}}, \bibinfo {author} {\bibfnamefont {B.~A.}\ \bibnamefont
  {Malomed}}, \bibinfo {author} {\bibfnamefont {L.}~\bibnamefont {Tarruell}}, \
  and\ \bibinfo {author} {\bibfnamefont {L.}~\bibnamefont {Torner}},\
  }\href@noop {} {\bibfield  {journal} {\bibinfo  {journal} {Phys. Rev. A}\
  }\textbf {\bibinfo {volume} {98}},\ \bibinfo {pages} {013612} (\bibinfo
  {year} {2018})}\BibitemShut {NoStop}%
\bibitem [{\citenamefont {Fort}\ and\ \citenamefont
  {Modugno}(2021)}]{fort2021self}%
  \BibitemOpen
  \bibfield  {author} {\bibinfo {author} {\bibfnamefont {C.}~\bibnamefont
  {Fort}}\ and\ \bibinfo {author} {\bibfnamefont {M.}~\bibnamefont {Modugno}},\
  }\href@noop {} {\bibfield  {journal} {\bibinfo  {journal} {Applied Sciences}\
  }\textbf {\bibinfo {volume} {11}},\ \bibinfo {pages} {866} (\bibinfo {year}
  {2021})}\BibitemShut {NoStop}%
\bibitem [{\citenamefont {Tolra}\ \emph {et~al.}(2004)\citenamefont {Tolra},
  \citenamefont {O’hara}, \citenamefont {Huckans}, \citenamefont {Phillips},
  \citenamefont {Rolston},\ and\ \citenamefont {Porto}}]{tolra2004observation}%
  \BibitemOpen
  \bibfield  {author} {\bibinfo {author} {\bibfnamefont {B.~L.}\ \bibnamefont
  {Tolra}}, \bibinfo {author} {\bibfnamefont {K.~M.}\ \bibnamefont {O’hara}},
  \bibinfo {author} {\bibfnamefont {J.~H.}\ \bibnamefont {Huckans}}, \bibinfo
  {author} {\bibfnamefont {W.~D.}\ \bibnamefont {Phillips}}, \bibinfo {author}
  {\bibfnamefont {S.~L.}\ \bibnamefont {Rolston}}, \ and\ \bibinfo {author}
  {\bibfnamefont {J.~V.}\ \bibnamefont {Porto}},\ }\href@noop {} {\bibfield
  {journal} {\bibinfo  {journal} {Phys. Rev. Lett.}\ }\textbf {\bibinfo
  {volume} {92}},\ \bibinfo {pages} {190401} (\bibinfo {year}
  {2004})}\BibitemShut {NoStop}%
\bibitem [{\citenamefont {Lavoine}\ and\ \citenamefont
  {Bourdel}(2021)}]{lavoine2021beyond}%
  \BibitemOpen
  \bibfield  {author} {\bibinfo {author} {\bibfnamefont {L.}~\bibnamefont
  {Lavoine}}\ and\ \bibinfo {author} {\bibfnamefont {T.}~\bibnamefont
  {Bourdel}},\ }\href@noop {} {\bibfield  {journal} {\bibinfo  {journal} {Phys.
  Rev. A}\ }\textbf {\bibinfo {volume} {103}},\ \bibinfo {pages} {033312}
  (\bibinfo {year} {2021})}\BibitemShut {NoStop}%
\bibitem [{\citenamefont {Ota}\ and\ \citenamefont
  {Astrakharchik}(2020)}]{ota2020beyond}%
  \BibitemOpen
  \bibfield  {author} {\bibinfo {author} {\bibfnamefont {M.}~\bibnamefont
  {Ota}}\ and\ \bibinfo {author} {\bibfnamefont {G.}~\bibnamefont
  {Astrakharchik}},\ }\href@noop {} {\bibfield  {journal} {\bibinfo  {journal}
  {SciPost Phys.}\ }\textbf {\bibinfo {volume} {9}},\ \bibinfo {pages} {1}
  (\bibinfo {year} {2020})}\BibitemShut {NoStop}%
\bibitem [{\citenamefont {Gu}\ and\ \citenamefont {Yin}(2020)}]{gu2020phonon}%
  \BibitemOpen
  \bibfield  {author} {\bibinfo {author} {\bibfnamefont {Q.}~\bibnamefont
  {Gu}}\ and\ \bibinfo {author} {\bibfnamefont {L.}~\bibnamefont {Yin}},\
  }\href@noop {} {\bibfield  {journal} {\bibinfo  {journal} {Phys. Rev. B}\
  }\textbf {\bibinfo {volume} {102}},\ \bibinfo {pages} {220503} (\bibinfo
  {year} {2020})}\BibitemShut {NoStop}%
\bibitem [{\citenamefont {Hu}\ and\ \citenamefont
  {Liu}(2020)}]{hu2020microscopic}%
  \BibitemOpen
  \bibfield  {author} {\bibinfo {author} {\bibfnamefont {H.}~\bibnamefont
  {Hu}}\ and\ \bibinfo {author} {\bibfnamefont {X.-J.}\ \bibnamefont {Liu}},\
  }\href@noop {} {\bibfield  {journal} {\bibinfo  {journal} {Phys. Rev. A}\
  }\textbf {\bibinfo {volume} {102}},\ \bibinfo {pages} {043302} (\bibinfo
  {year} {2020})}\BibitemShut {NoStop}%
\bibitem [{\citenamefont {Cikojevi{\'c}}\ \emph {et~al.}(2019)\citenamefont
  {Cikojevi{\'c}}, \citenamefont {Marki{\'c}}, \citenamefont {Astrakharchik},\
  and\ \citenamefont {Boronat}}]{cikojevic2019universality}%
  \BibitemOpen
  \bibfield  {author} {\bibinfo {author} {\bibfnamefont {V.}~\bibnamefont
  {Cikojevi{\'c}}}, \bibinfo {author} {\bibfnamefont {L.~V.}\ \bibnamefont
  {Marki{\'c}}}, \bibinfo {author} {\bibfnamefont {G.~E.}\ \bibnamefont
  {Astrakharchik}}, \ and\ \bibinfo {author} {\bibfnamefont {J.}~\bibnamefont
  {Boronat}},\ }\href@noop {} {\bibfield  {journal} {\bibinfo  {journal} {Phys.
  Rev. A}\ }\textbf {\bibinfo {volume} {99}},\ \bibinfo {pages} {023618}
  (\bibinfo {year} {2019})}\BibitemShut {NoStop}%
\bibitem [{\citenamefont {Parisi}\ and\ \citenamefont
  {Giorgini}(2020)}]{parisi2020quantum}%
  \BibitemOpen
  \bibfield  {author} {\bibinfo {author} {\bibfnamefont {L.}~\bibnamefont
  {Parisi}}\ and\ \bibinfo {author} {\bibfnamefont {S.}~\bibnamefont
  {Giorgini}},\ }\href@noop {} {\bibfield  {journal} {\bibinfo  {journal}
  {Phys. Rev. A}\ }\textbf {\bibinfo {volume} {102}},\ \bibinfo {pages}
  {023318} (\bibinfo {year} {2020})}\BibitemShut {NoStop}%
\bibitem [{\citenamefont {Parisi}\ \emph {et~al.}(2019)\citenamefont {Parisi},
  \citenamefont {Astrakharchik},\ and\ \citenamefont
  {Giorgini}}]{parisi2019liquid}%
  \BibitemOpen
  \bibfield  {author} {\bibinfo {author} {\bibfnamefont {L.}~\bibnamefont
  {Parisi}}, \bibinfo {author} {\bibfnamefont {G.~E.}\ \bibnamefont
  {Astrakharchik}}, \ and\ \bibinfo {author} {\bibfnamefont {S.}~\bibnamefont
  {Giorgini}},\ }\href@noop {} {\bibfield  {journal} {\bibinfo  {journal}
  {Phys. Rev. Lett.}\ }\textbf {\bibinfo {volume} {122}},\ \bibinfo {pages}
  {105302} (\bibinfo {year} {2019})}\BibitemShut {NoStop}%
\bibitem [{\citenamefont {Kinoshita}\ \emph {et~al.}(2006)\citenamefont
  {Kinoshita}, \citenamefont {Wenger},\ and\ \citenamefont
  {Weiss}}]{kinoshita2006quantum}%
  \BibitemOpen
  \bibfield  {author} {\bibinfo {author} {\bibfnamefont {T.}~\bibnamefont
  {Kinoshita}}, \bibinfo {author} {\bibfnamefont {T.}~\bibnamefont {Wenger}}, \
  and\ \bibinfo {author} {\bibfnamefont {D.~S.}\ \bibnamefont {Weiss}},\
  }\href@noop {} {\bibfield  {journal} {\bibinfo  {journal} {Nature}\ }\textbf
  {\bibinfo {volume} {440}},\ \bibinfo {pages} {900} (\bibinfo {year}
  {2006})}\BibitemShut {NoStop}%
\bibitem [{\citenamefont {Li}\ \emph {et~al.}(2020)\citenamefont {Li},
  \citenamefont {Zhou}, \citenamefont {Mazets}, \citenamefont {Stimming},
  \citenamefont {M{\o}ller}, \citenamefont {Zhu}, \citenamefont {Zhai},
  \citenamefont {Xiong}, \citenamefont {Zhou}, \citenamefont {Chen} \emph
  {et~al.}}]{li2020relaxation}%
  \BibitemOpen
  \bibfield  {author} {\bibinfo {author} {\bibfnamefont {C.}~\bibnamefont
  {Li}}, \bibinfo {author} {\bibfnamefont {T.}~\bibnamefont {Zhou}}, \bibinfo
  {author} {\bibfnamefont {I.}~\bibnamefont {Mazets}}, \bibinfo {author}
  {\bibfnamefont {H.-P.}\ \bibnamefont {Stimming}}, \bibinfo {author}
  {\bibfnamefont {F.~S.}\ \bibnamefont {M{\o}ller}}, \bibinfo {author}
  {\bibfnamefont {Z.}~\bibnamefont {Zhu}}, \bibinfo {author} {\bibfnamefont
  {Y.}~\bibnamefont {Zhai}}, \bibinfo {author} {\bibfnamefont {W.}~\bibnamefont
  {Xiong}}, \bibinfo {author} {\bibfnamefont {X.}~\bibnamefont {Zhou}},
  \bibinfo {author} {\bibfnamefont {X.}~\bibnamefont {Chen}},  \emph {et~al.},\
  }\href@noop {} {\bibfield  {journal} {\bibinfo  {journal} {SciPost Physics}\
  }\textbf {\bibinfo {volume} {9}},\ \bibinfo {pages} {058} (\bibinfo {year}
  {2020})}\BibitemShut {NoStop}%
\bibitem [{\citenamefont {Astrakharchik}\ and\ \citenamefont
  {Malomed}(2018)}]{astrakharchik2018dynamics}%
  \BibitemOpen
  \bibfield  {author} {\bibinfo {author} {\bibfnamefont {G.~E.}\ \bibnamefont
  {Astrakharchik}}\ and\ \bibinfo {author} {\bibfnamefont {B.~A.}\ \bibnamefont
  {Malomed}},\ }\href@noop {} {\bibfield  {journal} {\bibinfo  {journal} {Phys.
  Rev. A}\ }\textbf {\bibinfo {volume} {98}},\ \bibinfo {pages} {013631}
  (\bibinfo {year} {2018})}\BibitemShut {NoStop}%
\bibitem [{\citenamefont {Tylutki}\ \emph {et~al.}(2020)\citenamefont
  {Tylutki}, \citenamefont {Astrakharchik}, \citenamefont {Malomed},\ and\
  \citenamefont {Petrov}}]{tylutki2020collective}%
  \BibitemOpen
  \bibfield  {author} {\bibinfo {author} {\bibfnamefont {M.}~\bibnamefont
  {Tylutki}}, \bibinfo {author} {\bibfnamefont {G.~E.}\ \bibnamefont
  {Astrakharchik}}, \bibinfo {author} {\bibfnamefont {B.~A.}\ \bibnamefont
  {Malomed}}, \ and\ \bibinfo {author} {\bibfnamefont {D.~S.}\ \bibnamefont
  {Petrov}},\ }\href@noop {} {\bibfield  {journal} {\bibinfo  {journal} {Phys.
  Rev. A}\ }\textbf {\bibinfo {volume} {101}},\ \bibinfo {pages} {051601}
  (\bibinfo {year} {2020})}\BibitemShut {NoStop}%
\bibitem [{\citenamefont {De~Rosi}\ \emph {et~al.}(2021)\citenamefont
  {De~Rosi}, \citenamefont {Astrakharchik},\ and\ \citenamefont
  {Massignan}}]{de2021thermal}%
  \BibitemOpen
  \bibfield  {author} {\bibinfo {author} {\bibfnamefont {G.}~\bibnamefont
  {De~Rosi}}, \bibinfo {author} {\bibfnamefont {G.~E.}\ \bibnamefont
  {Astrakharchik}}, \ and\ \bibinfo {author} {\bibfnamefont {P.}~\bibnamefont
  {Massignan}},\ }\href@noop {} {\bibfield  {journal} {\bibinfo  {journal}
  {Phys. Rev. A}\ }\textbf {\bibinfo {volume} {103}},\ \bibinfo {pages}
  {043316} (\bibinfo {year} {2021})}\BibitemShut {NoStop}%
\bibitem [{\citenamefont {Wang}\ \emph {et~al.}(2020)\citenamefont {Wang},
  \citenamefont {Hu},\ and\ \citenamefont {Liu}}]{wang2020thermal}%
  \BibitemOpen
  \bibfield  {author} {\bibinfo {author} {\bibfnamefont {J.}~\bibnamefont
  {Wang}}, \bibinfo {author} {\bibfnamefont {H.}~\bibnamefont {Hu}}, \ and\
  \bibinfo {author} {\bibfnamefont {X.-J.}\ \bibnamefont {Liu}},\ }\href@noop
  {} {\bibfield  {journal} {\bibinfo  {journal} {New J. Phys.}\ }\textbf
  {\bibinfo {volume} {22}},\ \bibinfo {pages} {103044} (\bibinfo {year}
  {2020})}\BibitemShut {NoStop}%
\bibitem [{\citenamefont {Morera}\ \emph {et~al.}(2020)\citenamefont {Morera},
  \citenamefont {Astrakharchik}, \citenamefont {Polls},\ and\ \citenamefont
  {Juli{\'a}-D{\'\i}az}}]{morera2020quantum}%
  \BibitemOpen
  \bibfield  {author} {\bibinfo {author} {\bibfnamefont {I.}~\bibnamefont
  {Morera}}, \bibinfo {author} {\bibfnamefont {G.~E.}\ \bibnamefont
  {Astrakharchik}}, \bibinfo {author} {\bibfnamefont {A.}~\bibnamefont
  {Polls}}, \ and\ \bibinfo {author} {\bibfnamefont {B.}~\bibnamefont
  {Juli{\'a}-D{\'\i}az}},\ }\href@noop {} {\bibfield  {journal} {\bibinfo
  {journal} {Phys. Rev. Research}\ }\textbf {\bibinfo {volume} {2}},\ \bibinfo
  {pages} {022008} (\bibinfo {year} {2020})}\BibitemShut {NoStop}%
\bibitem [{\citenamefont {Morera}\ \emph {et~al.}(2021)\citenamefont {Morera},
  \citenamefont {Astrakharchik}, \citenamefont {Polls},\ and\ \citenamefont
  {Juli{\'a}-D{\'\i}az}}]{morera2021universal}%
  \BibitemOpen
  \bibfield  {author} {\bibinfo {author} {\bibfnamefont {I.}~\bibnamefont
  {Morera}}, \bibinfo {author} {\bibfnamefont {G.~E.}\ \bibnamefont
  {Astrakharchik}}, \bibinfo {author} {\bibfnamefont {A.}~\bibnamefont
  {Polls}}, \ and\ \bibinfo {author} {\bibfnamefont {B.}~\bibnamefont
  {Juli{\'a}-D{\'\i}az}},\ }\href@noop {} {\bibfield  {journal} {\bibinfo
  {journal} {Phys. Rev. Lett.}\ }\textbf {\bibinfo {volume} {126}},\ \bibinfo
  {pages} {023001} (\bibinfo {year} {2021})}\BibitemShut {NoStop}%
\bibitem [{\citenamefont {J{\o}rgensen}\ \emph {et~al.}(2018)\citenamefont
  {J{\o}rgensen}, \citenamefont {Bruun},\ and\ \citenamefont
  {Arlt}}]{jorgensen2018dilute}%
  \BibitemOpen
  \bibfield  {author} {\bibinfo {author} {\bibfnamefont {N.~B.}\ \bibnamefont
  {J{\o}rgensen}}, \bibinfo {author} {\bibfnamefont {G.~M.}\ \bibnamefont
  {Bruun}}, \ and\ \bibinfo {author} {\bibfnamefont {J.~J.}\ \bibnamefont
  {Arlt}},\ }\href@noop {} {\bibfield  {journal} {\bibinfo  {journal} {Phys.
  Rev. Lett.}\ }\textbf {\bibinfo {volume} {121}},\ \bibinfo {pages} {173403}
  (\bibinfo {year} {2018})}\BibitemShut {NoStop}%
\bibitem [{\citenamefont {Skov}\ \emph {et~al.}(2021)\citenamefont {Skov},
  \citenamefont {Skou}, \citenamefont {J{\o}rgensen},\ and\ \citenamefont
  {Arlt}}]{skov2021observation}%
  \BibitemOpen
  \bibfield  {author} {\bibinfo {author} {\bibfnamefont {T.~G.}\ \bibnamefont
  {Skov}}, \bibinfo {author} {\bibfnamefont {M.~G.}\ \bibnamefont {Skou}},
  \bibinfo {author} {\bibfnamefont {N.~B.}\ \bibnamefont {J{\o}rgensen}}, \
  and\ \bibinfo {author} {\bibfnamefont {J.~J.}\ \bibnamefont {Arlt}},\
  }\href@noop {} {\bibfield  {journal} {\bibinfo  {journal} {Phys. Rev. Lett.}\
  }\textbf {\bibinfo {volume} {126}},\ \bibinfo {pages} {230404} (\bibinfo
  {year} {2021})}\BibitemShut {NoStop}%
\bibitem [{\citenamefont {Guo}\ \emph {et~al.}(2021)\citenamefont {Guo},
  \citenamefont {Jia}, \citenamefont {Li}, \citenamefont {Ma}, \citenamefont
  {Hutson}, \citenamefont {Cui},\ and\ \citenamefont
  {Wang}}]{guo2021leehuangyang}%
  \BibitemOpen
  \bibfield  {author} {\bibinfo {author} {\bibfnamefont {Z.}~\bibnamefont
  {Guo}}, \bibinfo {author} {\bibfnamefont {F.}~\bibnamefont {Jia}}, \bibinfo
  {author} {\bibfnamefont {L.}~\bibnamefont {Li}}, \bibinfo {author}
  {\bibfnamefont {Y.}~\bibnamefont {Ma}}, \bibinfo {author} {\bibfnamefont
  {J.~M.}\ \bibnamefont {Hutson}}, \bibinfo {author} {\bibfnamefont
  {X.}~\bibnamefont {Cui}}, \ and\ \bibinfo {author} {\bibfnamefont
  {D.}~\bibnamefont {Wang}},\ }\href@noop {} {\  (\bibinfo {year} {2021})},\
  \Eprint {http://arxiv.org/abs/2105.01277} {arXiv:2105.01277
  [cond-mat.quant-gas]} \BibitemShut {NoStop}%
\bibitem [{\citenamefont {Liu}\ \emph {et~al.}(2019)\citenamefont {Liu},
  \citenamefont {Zhang}, \citenamefont {Zhong}, \citenamefont {Zhang},
  \citenamefont {Qin}, \citenamefont {Huang}, \citenamefont {Li},\ and\
  \citenamefont {Malomed}}]{liu2019symmetry}%
  \BibitemOpen
  \bibfield  {author} {\bibinfo {author} {\bibfnamefont {B.}~\bibnamefont
  {Liu}}, \bibinfo {author} {\bibfnamefont {H.-F.}\ \bibnamefont {Zhang}},
  \bibinfo {author} {\bibfnamefont {R.-X.}\ \bibnamefont {Zhong}}, \bibinfo
  {author} {\bibfnamefont {X.-L.}\ \bibnamefont {Zhang}}, \bibinfo {author}
  {\bibfnamefont {X.-Z.}\ \bibnamefont {Qin}}, \bibinfo {author} {\bibfnamefont
  {C.}~\bibnamefont {Huang}}, \bibinfo {author} {\bibfnamefont {Y.-Y.}\
  \bibnamefont {Li}}, \ and\ \bibinfo {author} {\bibfnamefont {B.~A.}\
  \bibnamefont {Malomed}},\ }\href@noop {} {\bibfield  {journal} {\bibinfo
  {journal} {Phys. Rev. A}\ }\textbf {\bibinfo {volume} {99}},\ \bibinfo
  {pages} {053602} (\bibinfo {year} {2019})}\BibitemShut {NoStop}%
\bibitem [{\citenamefont {Mithun}\ \emph {et~al.}(2020)\citenamefont {Mithun},
  \citenamefont {Maluckov}, \citenamefont {Kasamatsu}, \citenamefont
  {Malomed},\ and\ \citenamefont {Khare}}]{mithun2020modulational}%
  \BibitemOpen
  \bibfield  {author} {\bibinfo {author} {\bibfnamefont {T.}~\bibnamefont
  {Mithun}}, \bibinfo {author} {\bibfnamefont {A.}~\bibnamefont {Maluckov}},
  \bibinfo {author} {\bibfnamefont {K.}~\bibnamefont {Kasamatsu}}, \bibinfo
  {author} {\bibfnamefont {B.~A.}\ \bibnamefont {Malomed}}, \ and\ \bibinfo
  {author} {\bibfnamefont {A.}~\bibnamefont {Khare}},\ }\href@noop {}
  {\bibfield  {journal} {\bibinfo  {journal} {Symmetry}\ }\textbf {\bibinfo
  {volume} {12}},\ \bibinfo {pages} {174} (\bibinfo {year} {2020})}\BibitemShut
  {NoStop}%
\bibitem [{\citenamefont {Cao}\ \emph {et~al.}(2017)\citenamefont {Cao},
  \citenamefont {Bolsinger}, \citenamefont {Mistakidis}, \citenamefont
  {Koutentakis}, \citenamefont {Kr{\"o}nke}, \citenamefont {Schurer},\ and\
  \citenamefont {Schmelcher}}]{cao2017unified}%
  \BibitemOpen
  \bibfield  {author} {\bibinfo {author} {\bibfnamefont {L.}~\bibnamefont
  {Cao}}, \bibinfo {author} {\bibfnamefont {V.}~\bibnamefont {Bolsinger}},
  \bibinfo {author} {\bibfnamefont {S.~I.}\ \bibnamefont {Mistakidis}},
  \bibinfo {author} {\bibfnamefont {G.~M.}\ \bibnamefont {Koutentakis}},
  \bibinfo {author} {\bibfnamefont {S.}~\bibnamefont {Kr{\"o}nke}}, \bibinfo
  {author} {\bibfnamefont {J.~M.}\ \bibnamefont {Schurer}}, \ and\ \bibinfo
  {author} {\bibfnamefont {P.}~\bibnamefont {Schmelcher}},\ }\href@noop {}
  {\bibfield  {journal} {\bibinfo  {journal} {J. Chem. Phys.}\ }\textbf
  {\bibinfo {volume} {147}},\ \bibinfo {pages} {044106} (\bibinfo {year}
  {2017})}\BibitemShut {NoStop}%
\bibitem [{\citenamefont {Cao}\ \emph {et~al.}(2013)\citenamefont {Cao},
  \citenamefont {Kr{\"o}nke}, \citenamefont {Vendrell},\ and\ \citenamefont
  {Schmelcher}}]{cao2013multi}%
  \BibitemOpen
  \bibfield  {author} {\bibinfo {author} {\bibfnamefont {L.}~\bibnamefont
  {Cao}}, \bibinfo {author} {\bibfnamefont {S.}~\bibnamefont {Kr{\"o}nke}},
  \bibinfo {author} {\bibfnamefont {O.}~\bibnamefont {Vendrell}}, \ and\
  \bibinfo {author} {\bibfnamefont {P.}~\bibnamefont {Schmelcher}},\
  }\href@noop {} {\bibfield  {journal} {\bibinfo  {journal} {J. Chem. Phys.}\
  }\textbf {\bibinfo {volume} {139}},\ \bibinfo {pages} {134103} (\bibinfo
  {year} {2013})}\BibitemShut {NoStop}%
\bibitem [{\citenamefont {Mistakidis}\ \emph {et~al.}(2018)\citenamefont
  {Mistakidis}, \citenamefont {Katsimiga}, \citenamefont {Kevrekidis},\ and\
  \citenamefont {Schmelcher}}]{mistakidis2018correlation}%
  \BibitemOpen
  \bibfield  {author} {\bibinfo {author} {\bibfnamefont {S.~I.}\ \bibnamefont
  {Mistakidis}}, \bibinfo {author} {\bibfnamefont {G.~C.}\ \bibnamefont
  {Katsimiga}}, \bibinfo {author} {\bibfnamefont {P.~G.}\ \bibnamefont
  {Kevrekidis}}, \ and\ \bibinfo {author} {\bibfnamefont {P.}~\bibnamefont
  {Schmelcher}},\ }\href@noop {} {\bibfield  {journal} {\bibinfo  {journal}
  {New J. Phys.}\ }\textbf {\bibinfo {volume} {20}},\ \bibinfo {pages} {043052}
  (\bibinfo {year} {2018})}\BibitemShut {NoStop}%
\bibitem [{\citenamefont {Mistakidis}\ \emph {et~al.}(2021)\citenamefont
  {Mistakidis}, \citenamefont {Koutentakis}, \citenamefont {Grusdt},
  \citenamefont {Sadeghpour},\ and\ \citenamefont
  {Schmelcher}}]{mistakidis2021radiofrequency}%
  \BibitemOpen
  \bibfield  {author} {\bibinfo {author} {\bibfnamefont {S.~I.}\ \bibnamefont
  {Mistakidis}}, \bibinfo {author} {\bibfnamefont {G.~M.}\ \bibnamefont
  {Koutentakis}}, \bibinfo {author} {\bibfnamefont {F.}~\bibnamefont {Grusdt}},
  \bibinfo {author} {\bibfnamefont {H.~R.}\ \bibnamefont {Sadeghpour}}, \ and\
  \bibinfo {author} {\bibfnamefont {P.}~\bibnamefont {Schmelcher}},\
  }\href@noop {} {\bibfield  {journal} {\bibinfo  {journal} {New J. Phys.}\
  }\textbf {\bibinfo {volume} {23}},\ \bibinfo {pages} {043051} (\bibinfo
  {year} {2021})}\BibitemShut {NoStop}%
\bibitem [{\citenamefont {Mistakidis}\ \emph
  {et~al.}(2019{\natexlab{a}})\citenamefont {Mistakidis}, \citenamefont
  {Katsimiga}, \citenamefont {Koutentakis}, \citenamefont {Busch},\ and\
  \citenamefont {Schmelcher}}]{mistakidis2019quench}%
  \BibitemOpen
  \bibfield  {author} {\bibinfo {author} {\bibfnamefont {S.~I.}\ \bibnamefont
  {Mistakidis}}, \bibinfo {author} {\bibfnamefont {G.~C.}\ \bibnamefont
  {Katsimiga}}, \bibinfo {author} {\bibfnamefont {G.~M.}\ \bibnamefont
  {Koutentakis}}, \bibinfo {author} {\bibfnamefont {T.}~\bibnamefont {Busch}},
  \ and\ \bibinfo {author} {\bibfnamefont {P.}~\bibnamefont {Schmelcher}},\
  }\href@noop {} {\bibfield  {journal} {\bibinfo  {journal} {Phys. Rev. Lett.}\
  }\textbf {\bibinfo {volume} {122}},\ \bibinfo {pages} {183001} (\bibinfo
  {year} {2019}{\natexlab{a}})}\BibitemShut {NoStop}%
\bibitem [{\citenamefont {Rodr{\'\i}guez-L{\'o}pez}\ and\ \citenamefont
  {Castellanos}(2021)}]{rodriguez2021oscillating}%
  \BibitemOpen
  \bibfield  {author} {\bibinfo {author} {\bibfnamefont {O.~A.}\ \bibnamefont
  {Rodr{\'\i}guez-L{\'o}pez}}\ and\ \bibinfo {author} {\bibfnamefont
  {E.}~\bibnamefont {Castellanos}},\ }\href@noop {} {\bibfield  {journal}
  {\bibinfo  {journal} {J. Low Temp. Phys.}\ ,\ \bibinfo {pages} {1}} (\bibinfo
  {year} {2021})}\BibitemShut {NoStop}%
\bibitem [{\citenamefont {Semeghini}\ \emph {et~al.}(2018)\citenamefont
  {Semeghini}, \citenamefont {Ferioli}, \citenamefont {Masi}, \citenamefont
  {Mazzinghi}, \citenamefont {Wolswijk}, \citenamefont {Minardi}, \citenamefont
  {Modugno}, \citenamefont {Modugno}, \citenamefont {Inguscio},\ and\
  \citenamefont {Fattori}}]{semeghini2018self}%
  \BibitemOpen
  \bibfield  {author} {\bibinfo {author} {\bibfnamefont {G.}~\bibnamefont
  {Semeghini}}, \bibinfo {author} {\bibfnamefont {G.}~\bibnamefont {Ferioli}},
  \bibinfo {author} {\bibfnamefont {L.}~\bibnamefont {Masi}}, \bibinfo {author}
  {\bibfnamefont {C.}~\bibnamefont {Mazzinghi}}, \bibinfo {author}
  {\bibfnamefont {L.}~\bibnamefont {Wolswijk}}, \bibinfo {author}
  {\bibfnamefont {F.}~\bibnamefont {Minardi}}, \bibinfo {author} {\bibfnamefont
  {M.}~\bibnamefont {Modugno}}, \bibinfo {author} {\bibfnamefont
  {G.}~\bibnamefont {Modugno}}, \bibinfo {author} {\bibfnamefont
  {M.}~\bibnamefont {Inguscio}}, \ and\ \bibinfo {author} {\bibfnamefont
  {M.}~\bibnamefont {Fattori}},\ }\href@noop {} {\bibfield  {journal} {\bibinfo
   {journal} {Phys. Rev. Lett.}\ }\textbf {\bibinfo {volume} {120}},\ \bibinfo
  {pages} {235301} (\bibinfo {year} {2018})}\BibitemShut {NoStop}%
\bibitem [{\citenamefont {Olshanii}(1998)}]{olshanii1998atomic}%
  \BibitemOpen
  \bibfield  {author} {\bibinfo {author} {\bibfnamefont {M.}~\bibnamefont
  {Olshanii}},\ }\href@noop {} {\bibfield  {journal} {\bibinfo  {journal}
  {Phys. Rev. Lett.}\ }\textbf {\bibinfo {volume} {81}},\ \bibinfo {pages}
  {938} (\bibinfo {year} {1998})}\BibitemShut {NoStop}%
\bibitem [{\citenamefont {Chin}\ \emph {et~al.}(2010)\citenamefont {Chin},
  \citenamefont {Grimm}, \citenamefont {Julienne},\ and\ \citenamefont
  {Tiesinga}}]{chin2010feshbach}%
  \BibitemOpen
  \bibfield  {author} {\bibinfo {author} {\bibfnamefont {C.}~\bibnamefont
  {Chin}}, \bibinfo {author} {\bibfnamefont {R.}~\bibnamefont {Grimm}},
  \bibinfo {author} {\bibfnamefont {P.}~\bibnamefont {Julienne}}, \ and\
  \bibinfo {author} {\bibfnamefont {E.}~\bibnamefont {Tiesinga}},\ }\href@noop
  {} {\bibfield  {journal} {\bibinfo  {journal} {Rev. Mod. Phys.}\ }\textbf
  {\bibinfo {volume} {82}},\ \bibinfo {pages} {1225} (\bibinfo {year}
  {2010})}\BibitemShut {NoStop}%
\bibitem [{\citenamefont {K{\"o}hler}\ \emph {et~al.}(2006)\citenamefont
  {K{\"o}hler}, \citenamefont {G{\'o}ral},\ and\ \citenamefont
  {Julienne}}]{kohler2006production}%
  \BibitemOpen
  \bibfield  {author} {\bibinfo {author} {\bibfnamefont {T.}~\bibnamefont
  {K{\"o}hler}}, \bibinfo {author} {\bibfnamefont {K.}~\bibnamefont
  {G{\'o}ral}}, \ and\ \bibinfo {author} {\bibfnamefont {P.~S.}\ \bibnamefont
  {Julienne}},\ }\href@noop {} {\bibfield  {journal} {\bibinfo  {journal} {Rev.
  Mod. Phys.}\ }\textbf {\bibinfo {volume} {78}},\ \bibinfo {pages} {1311}
  (\bibinfo {year} {2006})}\BibitemShut {NoStop}%
\bibitem [{\citenamefont {Gaunt}\ \emph {et~al.}(2013)\citenamefont {Gaunt},
  \citenamefont {Schmidutz}, \citenamefont {Gotlibovych}, \citenamefont
  {Smith},\ and\ \citenamefont {Hadzibabic}}]{gaunt2013bose}%
  \BibitemOpen
  \bibfield  {author} {\bibinfo {author} {\bibfnamefont {A.~L.}\ \bibnamefont
  {Gaunt}}, \bibinfo {author} {\bibfnamefont {T.~F.}\ \bibnamefont
  {Schmidutz}}, \bibinfo {author} {\bibfnamefont {I.}~\bibnamefont
  {Gotlibovych}}, \bibinfo {author} {\bibfnamefont {R.~P.}\ \bibnamefont
  {Smith}}, \ and\ \bibinfo {author} {\bibfnamefont {Z.}~\bibnamefont
  {Hadzibabic}},\ }\href@noop {} {\bibfield  {journal} {\bibinfo  {journal}
  {Phys. Rev. Lett.}\ }\textbf {\bibinfo {volume} {110}},\ \bibinfo {pages}
  {200406} (\bibinfo {year} {2013})}\BibitemShut {NoStop}%
\bibitem [{\citenamefont {Pitaevskii}\ and\ \citenamefont
  {Stringari}(2016)}]{pitaevskii2016bose}%
  \BibitemOpen
  \bibfield  {author} {\bibinfo {author} {\bibfnamefont {L.}~\bibnamefont
  {Pitaevskii}}\ and\ \bibinfo {author} {\bibfnamefont {S.}~\bibnamefont
  {Stringari}},\ }\href@noop {} {\emph {\bibinfo {title} {Bose-Einstein
  condensation and superfluidity}}},\ Vol.\ \bibinfo {volume} {164}\ (\bibinfo
  {publisher} {Oxford University Press},\ \bibinfo {year} {2016})\BibitemShut
  {NoStop}%
\bibitem [{\citenamefont {Lode}\ \emph {et~al.}(2020)\citenamefont {Lode},
  \citenamefont {L{\'e}v{\^e}que}, \citenamefont {Madsen}, \citenamefont
  {Streltsov},\ and\ \citenamefont {Alon}}]{lode2020colloquium}%
  \BibitemOpen
  \bibfield  {author} {\bibinfo {author} {\bibfnamefont {A.~U.~J.}\
  \bibnamefont {Lode}}, \bibinfo {author} {\bibfnamefont {C.}~\bibnamefont
  {L{\'e}v{\^e}que}}, \bibinfo {author} {\bibfnamefont {L.~B.}\ \bibnamefont
  {Madsen}}, \bibinfo {author} {\bibfnamefont {A.~I.}\ \bibnamefont
  {Streltsov}}, \ and\ \bibinfo {author} {\bibfnamefont {O.~E.}\ \bibnamefont
  {Alon}},\ }\href@noop {} {\bibfield  {journal} {\bibinfo  {journal} {Rev.
  Mod. Phys.}\ }\textbf {\bibinfo {volume} {92}},\ \bibinfo {pages} {011001}
  (\bibinfo {year} {2020})}\BibitemShut {NoStop}%
\bibitem [{\citenamefont {Katsimiga}\ \emph {et~al.}(2017)\citenamefont
  {Katsimiga}, \citenamefont {Koutentakis}, \citenamefont {Mistakidis},
  \citenamefont {Kevrekidis},\ and\ \citenamefont
  {Schmelcher}}]{katsimiga2017dark}%
  \BibitemOpen
  \bibfield  {author} {\bibinfo {author} {\bibfnamefont {G.~C.}\ \bibnamefont
  {Katsimiga}}, \bibinfo {author} {\bibfnamefont {G.~M.}\ \bibnamefont
  {Koutentakis}}, \bibinfo {author} {\bibfnamefont {S.~I.}\ \bibnamefont
  {Mistakidis}}, \bibinfo {author} {\bibfnamefont {P.~G.}\ \bibnamefont
  {Kevrekidis}}, \ and\ \bibinfo {author} {\bibfnamefont {P.}~\bibnamefont
  {Schmelcher}},\ }\href@noop {} {\bibfield  {journal} {\bibinfo  {journal}
  {New J. Phys.}\ }\textbf {\bibinfo {volume} {19}},\ \bibinfo {pages} {073004}
  (\bibinfo {year} {2017})}\BibitemShut {NoStop}%
\bibitem [{\citenamefont {Horodecki}\ \emph {et~al.}(2009)\citenamefont
  {Horodecki}, \citenamefont {Horodecki}, \citenamefont {Horodecki},\ and\
  \citenamefont {Horodecki}}]{horodecki2009quantum}%
  \BibitemOpen
  \bibfield  {author} {\bibinfo {author} {\bibfnamefont {R.}~\bibnamefont
  {Horodecki}}, \bibinfo {author} {\bibfnamefont {P.}~\bibnamefont
  {Horodecki}}, \bibinfo {author} {\bibfnamefont {M.}~\bibnamefont
  {Horodecki}}, \ and\ \bibinfo {author} {\bibfnamefont {K.}~\bibnamefont
  {Horodecki}},\ }\href@noop {} {\bibfield  {journal} {\bibinfo  {journal}
  {Rev. Mod. Phys.}\ }\textbf {\bibinfo {volume} {81}},\ \bibinfo {pages} {865}
  (\bibinfo {year} {2009})}\BibitemShut {NoStop}%
\bibitem [{\citenamefont {Roncaglia}\ \emph {et~al.}(2014)\citenamefont
  {Roncaglia}, \citenamefont {Montorsi},\ and\ \citenamefont
  {Genovese}}]{roncaglia2014bipartite}%
  \BibitemOpen
  \bibfield  {author} {\bibinfo {author} {\bibfnamefont {M.}~\bibnamefont
  {Roncaglia}}, \bibinfo {author} {\bibfnamefont {A.}~\bibnamefont {Montorsi}},
  \ and\ \bibinfo {author} {\bibfnamefont {M.}~\bibnamefont {Genovese}},\
  }\href@noop {} {\bibfield  {journal} {\bibinfo  {journal} {Phys. Rev. A}\
  }\textbf {\bibinfo {volume} {90}},\ \bibinfo {pages} {062303} (\bibinfo
  {year} {2014})}\BibitemShut {NoStop}%
\bibitem [{\citenamefont {Mistakidis}\ \emph
  {et~al.}(2019{\natexlab{b}})\citenamefont {Mistakidis}, \citenamefont
  {Volosniev}, \citenamefont {Zinner},\ and\ \citenamefont
  {Schmelcher}}]{mistakidis2019effective}%
  \BibitemOpen
  \bibfield  {author} {\bibinfo {author} {\bibfnamefont {S.~I.}\ \bibnamefont
  {Mistakidis}}, \bibinfo {author} {\bibfnamefont {A.~G.}\ \bibnamefont
  {Volosniev}}, \bibinfo {author} {\bibfnamefont {N.~T.}\ \bibnamefont
  {Zinner}}, \ and\ \bibinfo {author} {\bibfnamefont {P.}~\bibnamefont
  {Schmelcher}},\ }\href@noop {} {\bibfield  {journal} {\bibinfo  {journal}
  {Phys. Rev. A}\ }\textbf {\bibinfo {volume} {100}},\ \bibinfo {pages}
  {013619} (\bibinfo {year} {2019}{\natexlab{b}})}\BibitemShut {NoStop}%
\bibitem [{\citenamefont {Kevrekidis}\ \emph {et~al.}(2007)\citenamefont
  {Kevrekidis}, \citenamefont {Frantzeskakis},\ and\ \citenamefont
  {Carretero-Gonz{\'a}lez}}]{kevrekidis2007emergent}%
  \BibitemOpen
  \bibfield  {author} {\bibinfo {author} {\bibfnamefont {P.~G.}\ \bibnamefont
  {Kevrekidis}}, \bibinfo {author} {\bibfnamefont {D.~J.}\ \bibnamefont
  {Frantzeskakis}}, \ and\ \bibinfo {author} {\bibfnamefont {R.}~\bibnamefont
  {Carretero-Gonz{\'a}lez}},\ }\href@noop {} {\emph {\bibinfo {title} {Emergent
  nonlinear phenomena in Bose-Einstein condensates: theory and experiment}}},\
  Vol.~\bibinfo {volume} {45}\ (\bibinfo  {publisher} {Springer Science \&
  Business Media},\ \bibinfo {year} {2007})\BibitemShut {NoStop}%
\bibitem [{\citenamefont {Frenkel}(1934)}]{frenkel1934wave}%
  \BibitemOpen
  \bibfield  {author} {\bibinfo {author} {\bibfnamefont {J.}~\bibnamefont
  {Frenkel}},\ }\href@noop {} {\enquote {\bibinfo {title} {Wave mechanics,
  clarendon},}\ } (\bibinfo {year} {1934})\BibitemShut {NoStop}%
\bibitem [{\citenamefont {Dirac}(1930)}]{dirac_1930}%
  \BibitemOpen
  \bibfield  {author} {\bibinfo {author} {\bibfnamefont {P.~A.~M.}\
  \bibnamefont {Dirac}},\ }\href {\doibase 10.1017/S0305004100016108}
  {\bibfield  {journal} {\bibinfo  {journal} {Mathematical Proceedings of the
  Cambridge Philosophical Society}\ }\textbf {\bibinfo {volume} {26}},\
  \bibinfo {pages} {376–385} (\bibinfo {year} {1930})}\BibitemShut {NoStop}%
\bibitem [{\citenamefont {Mithun}\ \emph {et~al.}(2021)\citenamefont {Mithun},
  \citenamefont {Mistakidis}, \citenamefont {Schmelcher},\ and\ \citenamefont
  {Kevrekidis}}]{mithun2021statistical}%
  \BibitemOpen
  \bibfield  {author} {\bibinfo {author} {\bibfnamefont {T.}~\bibnamefont
  {Mithun}}, \bibinfo {author} {\bibfnamefont {S.~I.}\ \bibnamefont
  {Mistakidis}}, \bibinfo {author} {\bibfnamefont {P.}~\bibnamefont
  {Schmelcher}}, \ and\ \bibinfo {author} {\bibfnamefont {P.~G.}\ \bibnamefont
  {Kevrekidis}},\ }\href@noop {} {\bibfield  {journal} {\bibinfo  {journal}
  {Phys. Rev. A 104, 033316}\ } (\bibinfo {year} {2021})}\BibitemShut {NoStop}%
\bibitem [{\citenamefont {Sakmann}\ \emph {et~al.}(2008)\citenamefont
  {Sakmann}, \citenamefont {Streltsov}, \citenamefont {Alon},\ and\
  \citenamefont {Cederbaum}}]{sakmann2008reduced}%
  \BibitemOpen
  \bibfield  {author} {\bibinfo {author} {\bibfnamefont {K.}~\bibnamefont
  {Sakmann}}, \bibinfo {author} {\bibfnamefont {A.~I.}\ \bibnamefont
  {Streltsov}}, \bibinfo {author} {\bibfnamefont {O.~E.}\ \bibnamefont {Alon}},
  \ and\ \bibinfo {author} {\bibfnamefont {L.~S.}\ \bibnamefont {Cederbaum}},\
  }\href@noop {} {\bibfield  {journal} {\bibinfo  {journal} {Phys. Rev. A}\
  }\textbf {\bibinfo {volume} {78}},\ \bibinfo {pages} {023615} (\bibinfo
  {year} {2008})}\BibitemShut {NoStop}%
\bibitem [{\citenamefont {Bloch}\ \emph {et~al.}(2008)\citenamefont {Bloch},
  \citenamefont {Dalibard},\ and\ \citenamefont {Zwerger}}]{bloch2008many}%
  \BibitemOpen
  \bibfield  {author} {\bibinfo {author} {\bibfnamefont {I.}~\bibnamefont
  {Bloch}}, \bibinfo {author} {\bibfnamefont {J.}~\bibnamefont {Dalibard}}, \
  and\ \bibinfo {author} {\bibfnamefont {W.}~\bibnamefont {Zwerger}},\
  }\href@noop {} {\bibfield  {journal} {\bibinfo  {journal} {Rev. Mod. Phys.}\
  }\textbf {\bibinfo {volume} {80}},\ \bibinfo {pages} {885} (\bibinfo {year}
  {2008})}\BibitemShut {NoStop}%
\bibitem [{\citenamefont {Mukherjee}\ \emph {et~al.}(2020)\citenamefont
  {Mukherjee}, \citenamefont {Mistakidis}, \citenamefont {Majumder},\ and\
  \citenamefont {Schmelcher}}]{mukherjee2020induced}%
  \BibitemOpen
  \bibfield  {author} {\bibinfo {author} {\bibfnamefont {K.}~\bibnamefont
  {Mukherjee}}, \bibinfo {author} {\bibfnamefont {S.~I.}\ \bibnamefont
  {Mistakidis}}, \bibinfo {author} {\bibfnamefont {S.}~\bibnamefont
  {Majumder}}, \ and\ \bibinfo {author} {\bibfnamefont {P.}~\bibnamefont
  {Schmelcher}},\ }\href@noop {} {\bibfield  {journal} {\bibinfo  {journal}
  {Phys. Rev. A}\ }\textbf {\bibinfo {volume} {102}},\ \bibinfo {pages}
  {053317} (\bibinfo {year} {2020})}\BibitemShut {NoStop}%
\bibitem [{\citenamefont {Keiler}\ \emph {et~al.}(2020)\citenamefont {Keiler},
  \citenamefont {Mistakidis},\ and\ \citenamefont
  {Schmelcher}}]{keiler2020doping}%
  \BibitemOpen
  \bibfield  {author} {\bibinfo {author} {\bibfnamefont {K.}~\bibnamefont
  {Keiler}}, \bibinfo {author} {\bibfnamefont {S.~I.}\ \bibnamefont
  {Mistakidis}}, \ and\ \bibinfo {author} {\bibfnamefont {P.}~\bibnamefont
  {Schmelcher}},\ }\href@noop {} {\bibfield  {journal} {\bibinfo  {journal}
  {New J. Phys.}\ }\textbf {\bibinfo {volume} {22}},\ \bibinfo {pages} {083003}
  (\bibinfo {year} {2020})}\BibitemShut {NoStop}%
\bibitem [{Note1()}]{Note1}%
  \BibitemOpen
  \bibinfo {note} {The mean-field interaction parameter $Ng$ increases for
  larger $N$ while keeping $g$ fixed, thus rendering the gas stronger
  interacting. This explains the more prominent deviation between the MB and
  the MF approaches, compare Figs.~\ref {fig:den_sym_varyg} (c),
  (d).}\BibitemShut {Stop}%
\bibitem [{Note2()}]{Note2}%
  \BibitemOpen
  \bibinfo {note} {The crossover behavior from a Gaussian-shaped to a FT and
  then to a homogeneous profile for a specific $\delta g /g$ and increasing
  atom number occurs independently of the finite size of the system, i.e. the
  length of the box potential. In particular the relevant particle number to
  realize this transition is smaller for decreasing $L$. For example, within
  MGP, if $L=100$ and $g=0.05$ then for $\delta g/g=0.02$ [$\delta g/g=0.2$] we
  achieve a FT when $N>2000$ [$N>100$] and a homogeneous distribution for
  $N>10^5$ [$N>5000$].}\BibitemShut {Stop}%
\bibitem [{\citenamefont {Wenz}\ \emph {et~al.}(2013)\citenamefont {Wenz},
  \citenamefont {Z{\"u}rn}, \citenamefont {Murmann}, \citenamefont {Brouzos},
  \citenamefont {Lompe},\ and\ \citenamefont {Jochim}}]{wenz2013few}%
  \BibitemOpen
  \bibfield  {author} {\bibinfo {author} {\bibfnamefont {A.~N.}\ \bibnamefont
  {Wenz}}, \bibinfo {author} {\bibfnamefont {G.}~\bibnamefont {Z{\"u}rn}},
  \bibinfo {author} {\bibfnamefont {S.}~\bibnamefont {Murmann}}, \bibinfo
  {author} {\bibfnamefont {I.}~\bibnamefont {Brouzos}}, \bibinfo {author}
  {\bibfnamefont {T.}~\bibnamefont {Lompe}}, \ and\ \bibinfo {author}
  {\bibfnamefont {S.}~\bibnamefont {Jochim}},\ }\href@noop {} {\bibfield
  {journal} {\bibinfo  {journal} {Science}\ }\textbf {\bibinfo {volume}
  {342}},\ \bibinfo {pages} {457} (\bibinfo {year} {2013})}\BibitemShut
  {NoStop}%
\bibitem [{\citenamefont {Katsimiga}\ \emph {et~al.}(2020)\citenamefont
  {Katsimiga}, \citenamefont {Mistakidis}, \citenamefont {Bersano},
  \citenamefont {Ome}, \citenamefont {Mossman}, \citenamefont {Mukherjee},
  \citenamefont {Schmelcher}, \citenamefont {Engels},\ and\ \citenamefont
  {Kevrekidis}}]{katsimiga2020observation}%
  \BibitemOpen
  \bibfield  {author} {\bibinfo {author} {\bibfnamefont {G.~C.}\ \bibnamefont
  {Katsimiga}}, \bibinfo {author} {\bibfnamefont {S.~I.}\ \bibnamefont
  {Mistakidis}}, \bibinfo {author} {\bibfnamefont {T.~M.}\ \bibnamefont
  {Bersano}}, \bibinfo {author} {\bibfnamefont {M.~K.~H.}\ \bibnamefont {Ome}},
  \bibinfo {author} {\bibfnamefont {S.~M.}\ \bibnamefont {Mossman}}, \bibinfo
  {author} {\bibfnamefont {K.}~\bibnamefont {Mukherjee}}, \bibinfo {author}
  {\bibfnamefont {P.}~\bibnamefont {Schmelcher}}, \bibinfo {author}
  {\bibfnamefont {P.}~\bibnamefont {Engels}}, \ and\ \bibinfo {author}
  {\bibfnamefont {P.~G.}\ \bibnamefont {Kevrekidis}},\ }\href@noop {}
  {\bibfield  {journal} {\bibinfo  {journal} {Phys. Rev. A}\ }\textbf {\bibinfo
  {volume} {102}},\ \bibinfo {pages} {023301} (\bibinfo {year}
  {2020})}\BibitemShut {NoStop}%
\bibitem [{\citenamefont {Cui}\ and\ \citenamefont
  {Ma}(2021)}]{cui2021droplet}%
  \BibitemOpen
  \bibfield  {author} {\bibinfo {author} {\bibfnamefont {X.}~\bibnamefont
  {Cui}}\ and\ \bibinfo {author} {\bibfnamefont {Y.}~\bibnamefont {Ma}},\
  }\href@noop {} {\bibfield  {journal} {\bibinfo  {journal} {Phys. Rev.
  Research}\ }\textbf {\bibinfo {volume} {3}},\ \bibinfo {pages} {L012027}
  (\bibinfo {year} {2021})}\BibitemShut {NoStop}%
\bibitem [{\citenamefont {Naidon}\ and\ \citenamefont
  {Petrov}(2021)}]{naidon2021mixed}%
  \BibitemOpen
  \bibfield  {author} {\bibinfo {author} {\bibfnamefont {P.}~\bibnamefont
  {Naidon}}\ and\ \bibinfo {author} {\bibfnamefont {D.~S.}\ \bibnamefont
  {Petrov}},\ }\href@noop {} {\bibfield  {journal} {\bibinfo  {journal} {Phys.
  Rev. Lett.}\ }\textbf {\bibinfo {volume} {126}},\ \bibinfo {pages} {115301}
  (\bibinfo {year} {2021})}\BibitemShut {NoStop}%
\bibitem [{\citenamefont {Minardi}\ \emph {et~al.}(2019)\citenamefont
  {Minardi}, \citenamefont {Ancilotto}, \citenamefont {Burchianti},
  \citenamefont {D'Errico}, \citenamefont {Fort},\ and\ \citenamefont
  {Modugno}}]{minardi2019effective}%
  \BibitemOpen
  \bibfield  {author} {\bibinfo {author} {\bibfnamefont {F.}~\bibnamefont
  {Minardi}}, \bibinfo {author} {\bibfnamefont {F.}~\bibnamefont {Ancilotto}},
  \bibinfo {author} {\bibfnamefont {A.}~\bibnamefont {Burchianti}}, \bibinfo
  {author} {\bibfnamefont {C.}~\bibnamefont {D'Errico}}, \bibinfo {author}
  {\bibfnamefont {C.}~\bibnamefont {Fort}}, \ and\ \bibinfo {author}
  {\bibfnamefont {M.}~\bibnamefont {Modugno}},\ }\href@noop {} {\bibfield
  {journal} {\bibinfo  {journal} {Phys. Rev. A}\ }\textbf {\bibinfo {volume}
  {100}},\ \bibinfo {pages} {063636} (\bibinfo {year} {2019})}\BibitemShut
  {NoStop}%
\bibitem [{\citenamefont {Maity}\ \emph {et~al.}(2020)\citenamefont {Maity},
  \citenamefont {Mukherjee}, \citenamefont {Mistakidis}, \citenamefont {Das},
  \citenamefont {Kevrekidis}, \citenamefont {Majumder},\ and\ \citenamefont
  {Schmelcher}}]{maity2020parametrically}%
  \BibitemOpen
  \bibfield  {author} {\bibinfo {author} {\bibfnamefont {D.~K.}\ \bibnamefont
  {Maity}}, \bibinfo {author} {\bibfnamefont {K.}~\bibnamefont {Mukherjee}},
  \bibinfo {author} {\bibfnamefont {S.~I.}\ \bibnamefont {Mistakidis}},
  \bibinfo {author} {\bibfnamefont {S.}~\bibnamefont {Das}}, \bibinfo {author}
  {\bibfnamefont {P.~G.}\ \bibnamefont {Kevrekidis}}, \bibinfo {author}
  {\bibfnamefont {S.}~\bibnamefont {Majumder}}, \ and\ \bibinfo {author}
  {\bibfnamefont {P.}~\bibnamefont {Schmelcher}},\ }\href@noop {} {\bibfield
  {journal} {\bibinfo  {journal} {Phys. Rev. A}\ }\textbf {\bibinfo {volume}
  {102}},\ \bibinfo {pages} {033320} (\bibinfo {year} {2020})}\BibitemShut
  {NoStop}%
\bibitem [{\citenamefont {Fukuhara}\ \emph {et~al.}(2013)\citenamefont
  {Fukuhara}, \citenamefont {Kantian}, \citenamefont {Endres}, \citenamefont
  {Cheneau}, \citenamefont {Schau{\ss}}, \citenamefont {Hild}, \citenamefont
  {Bellem}, \citenamefont {Schollw{\"o}ck}, \citenamefont {Giamarchi},
  \citenamefont {Gross} \emph {et~al.}}]{fukuhara2013quantum}%
  \BibitemOpen
  \bibfield  {author} {\bibinfo {author} {\bibfnamefont {T.}~\bibnamefont
  {Fukuhara}}, \bibinfo {author} {\bibfnamefont {A.}~\bibnamefont {Kantian}},
  \bibinfo {author} {\bibfnamefont {M.}~\bibnamefont {Endres}}, \bibinfo
  {author} {\bibfnamefont {M.}~\bibnamefont {Cheneau}}, \bibinfo {author}
  {\bibfnamefont {P.}~\bibnamefont {Schau{\ss}}}, \bibinfo {author}
  {\bibfnamefont {S.}~\bibnamefont {Hild}}, \bibinfo {author} {\bibfnamefont
  {D.}~\bibnamefont {Bellem}}, \bibinfo {author} {\bibfnamefont
  {U.}~\bibnamefont {Schollw{\"o}ck}}, \bibinfo {author} {\bibfnamefont
  {T.}~\bibnamefont {Giamarchi}}, \bibinfo {author} {\bibfnamefont
  {C.}~\bibnamefont {Gross}},  \emph {et~al.},\ }\href@noop {} {\bibfield
  {journal} {\bibinfo  {journal} {Nature Phys.}\ }\textbf {\bibinfo {volume}
  {9}},\ \bibinfo {pages} {235} (\bibinfo {year} {2013})}\BibitemShut {NoStop}%
\bibitem [{\citenamefont {Ronzheimer}\ \emph {et~al.}(2013)\citenamefont
  {Ronzheimer}, \citenamefont {Schreiber}, \citenamefont {Braun}, \citenamefont
  {Hodgman}, \citenamefont {Langer}, \citenamefont {McCulloch}, \citenamefont
  {Heidrich-Meisner}, \citenamefont {Bloch},\ and\ \citenamefont
  {Schneider}}]{ronzheimer2013expansion}%
  \BibitemOpen
  \bibfield  {author} {\bibinfo {author} {\bibfnamefont {J.~P.}\ \bibnamefont
  {Ronzheimer}}, \bibinfo {author} {\bibfnamefont {M.}~\bibnamefont
  {Schreiber}}, \bibinfo {author} {\bibfnamefont {S.}~\bibnamefont {Braun}},
  \bibinfo {author} {\bibfnamefont {S.~S.}\ \bibnamefont {Hodgman}}, \bibinfo
  {author} {\bibfnamefont {S.}~\bibnamefont {Langer}}, \bibinfo {author}
  {\bibfnamefont {I.~P.}\ \bibnamefont {McCulloch}}, \bibinfo {author}
  {\bibfnamefont {F.}~\bibnamefont {Heidrich-Meisner}}, \bibinfo {author}
  {\bibfnamefont {I.}~\bibnamefont {Bloch}}, \ and\ \bibinfo {author}
  {\bibfnamefont {U.}~\bibnamefont {Schneider}},\ }\href@noop {} {\bibfield
  {journal} {\bibinfo  {journal} {Phys. Rev. Lett.}\ }\textbf {\bibinfo
  {volume} {110}},\ \bibinfo {pages} {205301} (\bibinfo {year}
  {2013})}\BibitemShut {NoStop}%
\bibitem [{Note3()}]{Note3}%
  \BibitemOpen
  \bibinfo {note} {The breathing frequency for fixed $\delta g/g$ and $g$ shows
  a weakly decreasing tendency for a smaller particle number in accordance to
  the observations made in Ref.~\cite
  {parisi2020quantum,astrakharchik2018dynamics}.}\BibitemShut {Stop}%
\bibitem [{Note4()}]{Note4}%
  \BibitemOpen
  \bibinfo {note} {Notice that we do not observe any appreciable signatures of
  dephasing in the dynamics of $\mathinner {\langle {X^2(t)}\rangle }$ within
  the MGP approach at least for total evolution times $T \leq 20000$. This
  statement holds irrespectively of the postquench value e.g. $(\delta
  g/g)_{\protect \rm f}=0,0.01,0.05$ and particle numbers
  $N=10,20,40,100$.}\BibitemShut {Stop}%
\bibitem [{\citenamefont {Nguyen}\ \emph {et~al.}(2019)\citenamefont {Nguyen},
  \citenamefont {Tsatsos}, \citenamefont {Luo}, \citenamefont {Lode},
  \citenamefont {Telles}, \citenamefont {Bagnato},\ and\ \citenamefont
  {Hulet}}]{nguyen2019parametric}%
  \BibitemOpen
  \bibfield  {author} {\bibinfo {author} {\bibfnamefont {J.~H.~V.}\
  \bibnamefont {Nguyen}}, \bibinfo {author} {\bibfnamefont {M.~C.}\
  \bibnamefont {Tsatsos}}, \bibinfo {author} {\bibfnamefont {D.}~\bibnamefont
  {Luo}}, \bibinfo {author} {\bibfnamefont {A.~U.~J.}\ \bibnamefont {Lode}},
  \bibinfo {author} {\bibfnamefont {G.~D.}\ \bibnamefont {Telles}}, \bibinfo
  {author} {\bibfnamefont {V.~S.}\ \bibnamefont {Bagnato}}, \ and\ \bibinfo
  {author} {\bibfnamefont {R.~G.}\ \bibnamefont {Hulet}},\ }\href@noop {}
  {\bibfield  {journal} {\bibinfo  {journal} {Phys. Rev. X}\ }\textbf {\bibinfo
  {volume} {9}},\ \bibinfo {pages} {011052} (\bibinfo {year}
  {2019})}\BibitemShut {NoStop}%
\bibitem [{\citenamefont {Zhang}\ \emph {et~al.}(2019)\citenamefont {Zhang},
  \citenamefont {Xu}, \citenamefont {Zheng}, \citenamefont {Chen},
  \citenamefont {Liu}, \citenamefont {Huang}, \citenamefont {Malomed},\ and\
  \citenamefont {Li}}]{PhysRevLett.123.133901}%
  \BibitemOpen
  \bibfield  {author} {\bibinfo {author} {\bibfnamefont {X.}~\bibnamefont
  {Zhang}}, \bibinfo {author} {\bibfnamefont {X.}~\bibnamefont {Xu}}, \bibinfo
  {author} {\bibfnamefont {Y.}~\bibnamefont {Zheng}}, \bibinfo {author}
  {\bibfnamefont {Z.}~\bibnamefont {Chen}}, \bibinfo {author} {\bibfnamefont
  {B.}~\bibnamefont {Liu}}, \bibinfo {author} {\bibfnamefont {C.}~\bibnamefont
  {Huang}}, \bibinfo {author} {\bibfnamefont {B.~A.}\ \bibnamefont {Malomed}},
  \ and\ \bibinfo {author} {\bibfnamefont {Y.}~\bibnamefont {Li}},\ }\href
  {\doibase 10.1103/PhysRevLett.123.133901} {\bibfield  {journal} {\bibinfo
  {journal} {Phys. Rev. Lett.}\ }\textbf {\bibinfo {volume} {123}},\ \bibinfo
  {pages} {133901} (\bibinfo {year} {2019})}\BibitemShut {NoStop}%
\bibitem [{\citenamefont {Kartashov}\ \emph {et~al.}(2019)\citenamefont
  {Kartashov}, \citenamefont {Malomed},\ and\ \citenamefont
  {Torner}}]{PhysRevLett.122.193902}%
  \BibitemOpen
  \bibfield  {author} {\bibinfo {author} {\bibfnamefont {Y.~V.}\ \bibnamefont
  {Kartashov}}, \bibinfo {author} {\bibfnamefont {B.~A.}\ \bibnamefont
  {Malomed}}, \ and\ \bibinfo {author} {\bibfnamefont {L.}~\bibnamefont
  {Torner}},\ }\href {\doibase 10.1103/PhysRevLett.122.193902} {\bibfield
  {journal} {\bibinfo  {journal} {Phys. Rev. Lett.}\ }\textbf {\bibinfo
  {volume} {122}},\ \bibinfo {pages} {193902} (\bibinfo {year}
  {2019})}\BibitemShut {NoStop}%
\end{thebibliography}%
\bibliographystyle{apsrev4-1}

\end{document}